\documentclass[12pt]{article}
\usepackage{amsmath}
\usepackage{graphicx}
\usepackage{natbib}
\usepackage{url} 
\usepackage{ulem}
\usepackage{amssymb, amsthm, amstext, booktabs, float, multirow, subfigure, rotating}
\usepackage{enumitem}
\usepackage{color}
\usepackage{setspace}
\usepackage{authblk}

\numberwithin{equation}{section} 
\numberwithin{figure}{section} 
\numberwithin{table}{section}

\newtheorem{definition}{Definition}[section] 
\newtheorem{example}{Example}[section] 

\theoremstyle{remark} 
\newtheorem{remark}{Remark}[section]

\bibpunct{(}{)}{,}{a}{,}{,}

\def\ppn{\vskip 6pt \noindent }
\def\R{{\mathbb{R}}}

\def\P{{\mathbb{P}}}

\newcommand{{\Xs}}{{\cal X}}
\newcommand{{\Ys}}{{\cal Y}}
\newcommand{{\Ls}}{{\cal L}}
\newcommand{{\Ss}}{{\cal S}}
\newcommand{{\Ms}}{{\cal M}}
\newcommand{{\Gs}}{{\cal G}}
\newcommand{{\Hs}}{{\cal H}}
\newcommand{{\Ns}}{{\cal N}}
\newcommand{{\Is}}{{\cal I}}
\newcommand{{\As}}{{\cal A}}
\newcommand{{\Bs}}{{\cal B}}
\newcommand{{\Cs}}{{\cal C}}
\newcommand{{\Rs}}{{\cal R}}
\newcommand{{\Us}}{{\cal U}}
\newcommand{{\Es}}{{\cal E}}
\newcommand{{\Fs}}{{\cal F}}
\newcommand{{\pp}}{{\mathbf p}}
\newcommand{{\Ps}}{{\cal P}}
\newcommand{{\KK}}{{\mathbf K}}
\newcommand{{\HH}}{{\mathbf H}}
\newcommand{{\II}}{{\mathbf I}}
\newcommand{{\yy}}{{\mathbf y}}
\newcommand{{\ab}}{{\mathbf a}}

\newcommand*\diff{\mathop{}\!\mathrm{d}} 

\newcommand{{\toL}}{{\overset{\mathcal{L}}{\longrightarrow}\ }}

\newcommand{{\MC}}{{\,  *_{\text{\scalebox{0.65}{$\Ms$}}}\,  }}

\newcommand{{\dou}}{$\leadsto$\ }

\newcommand{\indic}[1]{
\hbox{${\it 1}\hskip -4.5pt I_{\{ #1 \}}$}
}

\title{Statistical depth in abstract metric spaces}
\author[1]{\textsc{Gery Geenens}\thanks{Corresponding author: {\tt ggeenens@unsw.edu.au}.}}
\author[2]{\textsc{Alicia Nieto-Reyes}\thanks{{\tt alicia.nieto@unican.es}.}}
\author[2,3]{\textsc{Giacomo Francisci}\thanks{{\tt giacomo.francisci@unitn.it}.}}
\affil[1]{School of Mathematics and Statistics, UNSW Sydney, Australia}
\affil[2]{Department of Mathematics, Statistics and Computer Science\\  \ University of Cantabria, Spain}
\affil[3]{Department of Mathematics, University of Trento, Italy}

\begin{document}

\def\spacingset#1{\renewcommand{\baselinestretch}%
{#1}\small\normalsize} \spacingset{1}


%
%
%
%

  \maketitle

\bigskip
\begin{abstract}
	The concept of depth has proved very important for multivariate and functional data analysis, as it essentially acts as a surrogate for the notion a ranking of observations which is absent in more than one dimension. Motivated by the rapid development of technology, in particular the advent of `Big Data', we extend here that concept to general metric spaces, propose a natural depth measure and explore its properties as a statistical depth function. Working in a general metric space allows the depth to be tailored to the data at hand and to the ultimate goal of the analysis, a very desirable property given the polymorphic nature of modern data sets. This flexibility is thoroughly illustrated by several real data analyses.
\end{abstract}

\noindent%
\vfill

\spacingset{1.5} 
\section{Introduction}
\label{sec:intro}

Huge parts of statistical theory, especially its nonparametric side, heavily rely on the notion of ranks, see for instance \cite{Gibbons10}. However, ranks are not well defined in a multivariate framework as there exists no natural ordering in more than one dimension. This fact motivated \cite{Tukey75} to introduce the notion of {\it statistical depth} as a surrogate for `multivariate ranks'. Concretely, a depth is a measure of how central (or how outlying) a given point is with respect to a multivariate probability distribution. \cite{Zhuo00}, following some earlier considerations in \cite{Liu90}, formulated the properties that a valid depth measure should satisfy. Since then, depth-based procedures have proved very important tools for robust multivariate statistical analyses, e.g.\ see \cite{Liu99} and \cite{Li04,Li08}. \cite{Serfling06} and \cite{Mosler13} offer excellent short reviews of the ideas surrounding the concept of depth, while \cite{Hallin21} recently shed new light on the problem of `multivariate ranks'. 

\ppn The early 21st century has also seen such technological progress in recording devices and memory capacity, that any spatio-temporal phenomenon can now be recorded essentially in continuous time or space, giving rise to `functional' random objects. As a result, a solid theory for Functional Data Analysis (FDA) has been developed as well, allowing the extension of most of the classical problems of statistical inference from the multivariate context to the inherently infinite-dimensional functional case. In particular, functional versions of statistical depth have been investigated \citep{Fraiman01,Cuevas07,Lopez09,Dutta11,Lopez11,Sguera13,Chakraborty14,Hlubinka15,Nieto21,Nieto21b}. It is worth noting that an infinite-dimensional environment implies specific theoretical and practical challenges, making the extension from `multivariate' to `functional' a non-trivial one \citep{Nieto15}.

\ppn In this paper, we carry on with this gradual extension process by defining the statistical depth for complex random objects living in abstract metric spaces. 
Again, this extension is motivated by the rapid development of technology. Indeed, this is the `{\it Big Data}' era, in which digital data is recorded everywhere, all the time. The information that this huge amount of data contain may enable next-generation scientific breakthroughs, drive business forward or hold governments accountable. However, this is conditional on the existence of a statistical toolbox suitable for such Big Data, the profusion and nature of which inducing commensurate challenges. Indeed those data consist of objects as various as high-dimensional/infinite-dimensional vectors, matrices or functions representing images, shapes, movies, texts, handwriting or speech (to cite a few); and live streaming series thereof -- this is often summarised as `3V' ({\it Volume}, {\it Variety} and {\it Velocity}).   

\ppn Mainstream statistical techniques often fall short for analysing such complex mathematical objects. Yet, it remains true that any statistical analysis requires a sense of how close two instances of the object of interest are to one another. It is then only natural to assume that they live in a space where distances can be defined -- that is, in a certain {\it metric space} \citep{Snasel17}. This motivates the need for a statistical depth defined in an abstract metric space; hence our proposal of a `{\it metric depth}'.

\ppn The idea that the concept of multivariate statistical depth could be extended to general non-Euclidean settings can be traced back to \citet[Section 3.1]{Carrizosa96}. Later, \cite{Li11} were considering a depth-based procedure for analysing abundance data, which are typically high-dimensional discrete data with many observed 0's. Because of that particular structure, the classical Euclidean distance is not optimal for quantifying (dis)similarities between observations, and analysts in the field usually prefer more specific metrics such as the Bray-Curtis distance\footnote{The Bray-Curtis `distance' does not satisfy triangle inequality, hence it is rather a semi-distance.} \citep{Bray57}. In consequence, inspired by earlier works by \cite{Maa96} and \cite{Bartoszynski97}, \cite{Li11} devised a depth measure which could allow the proximity between observations to be quantified by a specific, user-chosen distance/dissimilarity measure.

\ppn This flexibility appears even more desirable when dealing with the polymorphous objects commonly found in modern data sets, as described above. 
For instance, functional objects are much richer than just infinite-dimensional vectors, and they can be compared on many different grounds: general appearance, short- or long-range variation, oscillating behaviour, etc.; which makes the choice of the `proximity measure' between two such objects a very crucial one \citep[Chapter 3]{Ferraty06}. On a more theoretical basis, an appropriate choice of such `proximity measure' sometimes allows one to get around issues caused by the `Curse of Dimensionality' \citep{Geenens11a}. 

\ppn Quantifying (dis)similarities between non-numeric objects is even more subject to discretionary choice. As an example, for comparing pieces of texts, the literature in text mining, linguistics and natural language processing proposed numerous metrics such as the Levenshtein distance, the Hamming distance, the Jaccard index or the Dice coefficient -- each targetting different dimensions of words, sentences or texts, such as similarity in spelling or similarity in meaning \citep{Wang20}. It is, therefore, paramount to have access to statistical procedures which allow a free choice of metric, and may be tailored to the kind of data at hand and to the ultimate purpose of the analysis.

\ppn Indeed, our proposed `metric depth' ($\mu D$), defined in Section \ref{subsec:FDBD}, enables such flexible analyses. Its main properties are explored in Section \ref{subsec:com} and an empirical version (computable from a sample) is described in Section \ref{sec:empdepth}. Section \ref{sec:realdat} illustrates its capabilities on several real data sets, including an application in `text mining' (Section \ref{sec:text}). Section \ref{sec:ccl} concludes.


\section{Statistical depth in metric spaces: definition}\label{subsec:FDBD}



Assume that the random object of interest, say $\Xs$, lives in a certain space $\Ms$ which can be equipped with a distance $d$. To avoid dispensable technical complications, it will be assumed throughout that $(\Ms,d)$ is a {\it complete} and {\it separable} metric space. Let $\As$ be the $\sigma$-algebra on $\Ms$ generated by the open $d$-metric balls and $\Ps$ be the space of all probability measures defined on the Borel sets of $\As$.\footnote{In a separable metric space, the Borel $\sigma$-algebra is generated by the open balls.} This makes $(\Ms,\As,P)$ a proper probability space for any $P \in \Ps$. In particular, it will be assumed that the distribution of $\Xs$ belongs to $\Ps$. 
Note that the cartesian product space $(\Ms \times \Ms, \As \times \As, P \times P)$ is then also a valid probability space \citep[Theorem I.1.10]{Parthasarathy67}. We denote:
\[\P(\Ss(\Xs_1,\Xs_2)) \doteq \iint_{\Ms \times \Ms} \Ss(\chi_1,\chi_2)\, \diff(P \times P)(\chi_1,\chi_2)\]
for any measurable statement $\Ss: \Ms \times \Ms \to \{0,1\}$ -- the statement returns the value 1 if it is true, and 0 otherwise. So, $\P(\Ss(\Xs_1,\Xs_2))$ returns the probability that $\Ss$ is true if $\Xs_1, \Xs_2$ are two independent replications of $\Xs$, whose distribution is $P$.


\ppn Then we give the following definition: 
\begin{definition} The `metric depth' (`$\mu D$') of the point $\chi$ in the metric space $(\Ms,d)$ with respect to the probability measure $P \in \Ps$ is defined as:
\begin{equation} 
 \mu D(\chi,P) =  \P\big( d(\Xs_1,\Xs_2) > \max\left\{d(\Xs_1,\chi),d(\Xs_2,\chi) \right\} \big). \label{eqn:FDBD}
\end{equation}
\end{definition}

\ppn For each fixed $\chi \in \Ms$, the set $\big\{(\chi_1,\chi_2)\in \Ms\times\Ms: d(\chi_1,\chi_2) > \max\{d(\chi_1,\chi),d(\chi_2,\chi)\} \big\}$ belongs to the $\sigma$-algebra $\As\times\As,$ with $\As$ defined above, making the probability statement $\P$ in (\ref{eqn:FDBD}) a well-defined one for any $P \in \Ps$. 

\ppn The interpretation of (\ref{eqn:FDBD}) in terms of depth is clear: a point $\chi \in \Ms$ is deep with respect to the distribution $P$ if it is likely to find it `between' two objects $\Xs_1$ and $\Xs_2$ in $\Ms$ randomly generated from $P$. `Between' here means that the side joining $\Xs_1$ and $\Xs_2$ is the longest in a `triangle' of $\Ms$ with vertices $\Xs_1$, $\Xs_2$ and $\chi$, or, in other words, that $\chi$ belongs to the intersection of the two open $d$-balls $B_d(\Xs_1,d(\Xs_1,\Xs_2))$ and $B_d(\Xs_2,d(\Xs_1,\Xs_2))$, 
where $B_d(\Xs_1,d(\Xs_1,\Xs_2))$ is the ball with center $\Xs_1$ and radius $d(\Xs_1,\Xs_2)$. 
In this sense, (\ref{eqn:FDBD}) is an extension of the vectorial `lens depth' \citep{Liu11}. If we define 
\begin{equation} L_d(\Xs_1,\Xs_2) := B_d(\Xs_1,d(\Xs_1,\Xs_2)) \cap B_d(\Xs_2,d(\Xs_1,\Xs_2)), \label{eqn:L} \end{equation}
 the `lens' defined by $\Xs_1$ and $\Xs_2$ in $(\Ms,d)$, then $\mu D(\chi,P) = \P( L_d(\Xs_1,\Xs_2) \ni \chi)$. This is the probability that a random set contains a certain element $\chi$, and interesting parallels can be drawn with the theory of random sets, in particular Choquet capacities and related ideas \citep[Chapter 1]{Molchanov05}. Note that, independently of this work, \cite{Cholaquidis20} recently explored the extension of the `lens depth' to general metric spaces as well. Their focus and the content of their paper are, however, much different to what is investigated here. 

\section{Main properties} \label{subsec:com}
The fact that the distance $d$ is left free
really makes the metric depth $\mu D$ a very flexible tool, as any meaningful $d$ equipping $\Ms$ can be used in (\ref{eqn:FDBD}) without altering the theoretical properties which we explore below.

\ppn In addition, we note that no-where in the developments, it is used explicitly the fact that $d(\chi,\xi) = 0 \iff \chi = \xi$ for any two $\chi,\xi \in \Ms$ (identity of indiscernibles). A proximity measure which satisfies all the properties of a distance (non-negativity, symmetry and triangle inequality) but not `identity of indiscernibles' is called a pseudo-distance. Hence, the metric depth (\ref{eqn:FDBD}) can be used in conjunction with a  pseudo-distance, while keeping its essential features. We can, for instance, assess the proximity between two objects by comparing the coefficients of their leading terms when expanded in certain bases, such as a spline basis in the case of functional data when smoothing the original data is necessary \citep[Chapter 3]{Ramsay05}. Other examples are given in Section \ref{sec:realdat}.


\subsection{Elasticity invariance}
\begin{enumerate} \item[$(P_1)$]
 Let $\varphi: \Ms  \to \Ms$ be an `elastic' map in the sense that for any $\chi,\xi, \chi',\xi' \in \Ms$, $d(\chi,\xi) < d(\chi',\xi') \iff d(\varphi(\chi),\varphi(\xi)) < d(\varphi(\chi'),\varphi(\xi'))$. Then, 
$\mu D(\varphi(\chi),P_{\varphi}) = \mu D(\chi,P),$ 
where $P_\varphi$ is the push-forward distribution of the image through $\varphi$ of a random object of $\Ms$ having distribution $P$.
\end{enumerate}

\ppn This follows from the fact that $d(\varphi(\Xs_1),\varphi(\Xs_2)) > \max\left\{d(\varphi(\Xs_1),\varphi(\chi)),d(\varphi(\Xs_2),\varphi(\chi)) \right\}$ $\iff d(\Xs_1,\Xs_2) > \max\left\{d(\Xs_1,\chi),d(\Xs_2,\chi) \right\}$ for such a map $\varphi$. These obviously include any isometry, such that $d(\chi,\xi) = d(\varphi(\chi),\varphi(\xi))$, or other dilation-type transformations such that  $d(\chi,\xi) = a_\varphi d(\varphi(\chi),\varphi(\xi))$, for some positive scalar constant $a_\varphi$, but not only. Clearly, $(P_1)$ establishes $\mu D$ as a purely topological concept. On another note, ($P_1$) may be thought of as an extension of property P1 in \citet[p.\,463]{Zhuo00} -- that a depth measure in $\R^d$ `{\it should not depend on the underlying coordinate system or, in particular, on the scales of the underlying measurements}'.

\subsection{Vanishing at infinity}

Assume that $(\Ms,d)$ is an unbounded metric space, i.e., $\sup_{\chi,\xi \in \Ms} d(\chi,\xi) = \infty$. Then:
\begin{enumerate} \item[$(P_2)$] For any $P \in \Ps$ and $\chi \in \Ms$, $\lim_{R \to \infty} \sup_{\xi \notin B_d(\chi,R) } \mu D(\xi,P) = 0$. \end{enumerate}
\ppn This follows from Proposition 1(a) in \cite{Cholaquidis20}. It is obviously the analogue to \cite{Zhuo00}'s P4: `{\it The depth of a point $x$ should approach 0 as $\|x\|$ approaches infinity}'.

\ppn Now suppose that, $\forall \chi \in \Ms$,
\begin{equation} \P\big( d(\Xs_1,\Xs_2) = \max\left\{d(\Xs_1,\chi),d(\Xs_2,\chi) \right\} \big) =0. \label{eqn:cont} \end{equation}
This kind of continuity condition guarantees that, with probability 1, a given $\chi \in \Ms$ will not lie exactly on the boundary of a random lens such as (\ref{eqn:L}). Then, we can prove the following properties $(P_3)$ and $(P_4)$.

\subsection{Continuity in $\chi$}

\begin{enumerate} \item[$(P_3)$]
	 For any $P \in \Ps$ such that (\ref{eqn:cont}) holds, $\forall \chi \in \Ms$ and $\forall \epsilon >0$, there exists $\delta >0$ such that \label{prop:P4} $$\sup_{\xi: d(\chi,\xi)<\delta} |\mu D(\xi,P) - \mu D(\chi,P)| < \epsilon.$$
\end{enumerate}
Indeed, for any $\chi \in \Ms$, take $\xi \in \Ms$ such that $d(\chi,\xi) < \delta$, for some $\delta>0$. Then, by the triangle inequality, for any $\Xs_1,\Xs_2 \in \Ms$, 
\[\max\left\{d(\Xs_1,\chi),d(\Xs_2,\chi) \right\} - \delta < \max\left\{d(\Xs_1,\xi),d(\Xs_2,\xi) \right\}< \max\left\{d(\Xs_1,\chi),d(\Xs_2,\chi) \right\} + \delta. \]
Hence, $\mu D(\xi,P)  = \P\big(  d(\Xs_1,\Xs_2) >  \max\left\{d(\Xs_1,\xi),d(\Xs_2,\xi) \right\} \big)$ is such that
\begin{multline*} \P\big(  d(\Xs_1,\Xs_2) >  \max\left\{d(\Xs_1,\chi),d(\Xs_2,\chi) \right\} + \delta \big) \\ \leq  \mu D(\xi,P) \leq  \P\big(  d(\Xs_1,\Xs_2) >  \max\left\{d(\Xs_1,\chi),d(\Xs_2,\chi) \right\} -\delta \big). \end{multline*}
Now, see that $\Psi(x) \doteq \P\left(   \max\left\{d(\Xs_1,\chi),d(\Xs_2,\chi) \right\} - d(\Xs_1,\Xs_2) \leq x \right)$ is a cumulative distribution function assumed to be continuous at $x=0$ by (\ref{eqn:cont}). This means that, for any $\epsilon>0$, we can find a $\delta >0$ such that $|\Psi(|\delta|)-\Psi(0)|< \epsilon$.  As $\Psi(0) = \mu D(\chi,P)$, the claim follows.

\subsection{Continuity in $P$}
\begin{enumerate} \item[$(P_4)$]
 For any $P \in \Ps$ such that (\ref{eqn:cont}) holds, $\forall \chi \in \Ms$ and $\forall \epsilon >0$, there exists $\delta >0$ such that $|\mu D(\chi,Q) - \mu D(\chi,P)| < \epsilon$ $P$-almost surely for all $Q \in \Ps$ with $d_\Ps(P,Q) <\delta$ $P$-almost surely, where $d_\Ps$ metricises the topology of weak convergence on $\Ps$. 
\end{enumerate}

\ppn This follows directly from classical results on convergence of probability measures on separable metric spaces -- e.g.\ \citet[Theorem 11.1.1]{Dudley02} -- as $\mu D(\chi,P)$ and $\mu D(\chi,Q)$ in (\ref{eqn:FDBD}) are simple probability statements on elements of $\As \times \As$. Note that (\ref{eqn:cont}) guarantees that the `lens' (\ref{eqn:L}) is a continuity set in the sense of \citet[Section 11.1]{Dudley02}.

\subsection{Further comments}

\cite{Zhuo00} listed two more desirable properties for a depth measure on $\R^d$: `Maximality at centre' and `Monotonicity relative to deepest point' (their properties P2 and P3). Similar features are difficult to investigate here for $\mu D$ without giving a stronger structure to $(\Ms,d)$, such as some sort of convexity, or $d$ to satisfy a parallelogram inequality, for example. As illustration, \cite{Zhuo00}'s P2 `Maximality at centre' requires the depth to be maximum at a uniquely defined `centre' with respect to some notion of symmetry. Without assuming a stronger structure $(\Ms,d)$, even the very definition of symmetry in $\Ms$ is unclear. As our aim here is to stay as flexible as possible with the proposed metric depth, we do not investigate further in that direction. Those properties of $\mu D$ may (or may not) be established on specific applications when $\Ms$ and $d$ are precisely defined, though.  

\ppn On a side note, even if \citet{Liu11} supposedly showed (their Theorem 6) that their Euclidean `lens depth' in $\R^d$ -- of which (\ref{eqn:FDBD}) can be thought of as an extension -- satisfies `Maximality at centre' for centrally symmetric distributions, their proof appears wrong as pointed out in \cite{Kleindessner17}. Yet,  \cite{Kleindessner17} conceded that they believe that the statement is true. In Appendix, we give three counter-examples, establishing that the statement is actually not true: \citet{Liu11}'s `lens depth' does not generally satisfy neither `Maximality at centre' nor `Monotonicity relative to deepest point' (\cite{Zhuo00}'s P2 and P3) for centrally symmetric distributions on $\R^d$.\footnote{\cite{Kleindessner17} noticed that analogous proofs for the spherical depth \citep{Elmore06} and the $\beta$-skeleton depth \citep{Yang17} are mistaken as well. Furthermore, we have found that the proof of a similar property for the band depth given in \citet[Theorem 1(2)]{Lopez09} is likewise erroneous.}

\ppn A last important point is the following. Suppose that the balls $B_d(\cdot,\cdot)$ are convex in $(\Ms,d)$. Then it can easily be checked that, for any non-degenerate distribution $P \in \Ps$ (i.e., not a unit point mass at some $\chi \in \Ms$), $\mu D$ cannot be degenerated in the sense that $\mu D(\chi,P) \equiv 0$ for all $\chi \in \Ms$.  Indeed, by convexity, the intersection $B_d(\Xs_1,d(\Xs_1,\Xs_2)) \cap B_d(\Xs_2,d(\Xs_1,\Xs_2))$ is non-empty as soon as $\Xs_1 \neq \Xs_2$, so, there always exists some $\chi \in \Ms$ which gets a positive depth by (\ref{eqn:FDBD}). It is known that some instances of statistical depth admit such a degenerate behaviour. For instance, that is the case of  \cite{Lopez09,Lopez11}'s band and half-region depths for a wide class of distributions on common functional spaces \citep[Theorems 3 and 4]{Chakraborty14}. 

\section{Empirical metric depth} \label{sec:empdepth}

Assume now that we have a random sample of realisations $\{\chi_i; i = 1 ,\ldots,n \}$ of the object $\Xs \sim P$ in $\Ms$. Then the depth of some point $\chi \in \Ms$ with respect to $P$ must actually be estimated. The empirical analogue of (\ref{eqn:FDBD}) is naturally $\mu D(\chi,\widehat{P}_n)$, where $\widehat{P}_n$ is the empirical measure of the sample, i.e., the collection of $1/n$-weighted point masses at the observed $\chi_1,\ldots,\chi_n$. This yields
\begin{equation}
 \mu D(\chi,\widehat{P}_n) = \frac{1}{\binom{n}{2}} \sum_{i < j} \indic{d(\chi_i,\chi_j) > \max\{d(\chi_i,\chi),d(\chi_j,\chi) \}}. \label{eqn:empDBFD}
\end{equation}
Obviously, $\widehat{P}_n  \overset{P\text{-a.s.}}{\longrightarrow} P$, which guarantees under (\ref{eqn:cont}) the strong pointwise consistency of the estimator $D(\chi,\widehat{P}_n)$, that is
\begin{equation} \mu D(\chi,\widehat{P}_n) \overset{P\text{-a.s.}}{\longrightarrow} \mu D(\chi,P), \label{eqn:pointconsist} \end{equation}
for all $\chi \in \Ms$. This easily follows from Property $(P_4).$ 

\begin{remark} {\it Universal uniform strong consistency of $\mu D$ on every subset of $\Ms$ that is equicontinuous with respect to $d$}: a desirable and much stronger result that (\ref{eqn:pointconsist}) is the universal uniform strong consistency of the depth measure $\mu D$ on a subset $\Phi$ of $\Ms$. This is defined as
\[ \sup_{\chi \in \Phi} |\mu D(\chi,\widehat{P}_n) - \mu D(\chi,P)| \overset{\text{a.s}}{\longrightarrow} 0, \qquad n \to \infty,\]
for any $P \in \Ps$. Note that, from (\ref{eqn:L}), (\ref{eqn:empDBFD}) can also be written
\begin{equation} \mu D(\chi,\widehat{P}_n) = \frac{1}{\binom{n}{2}} \sum_{i < j} \indic{\chi \in B_d(\chi_i,d(\chi_i,\chi_j)) \cap B_d(\chi_j,d(\chi_i,\chi_j))}. \label{eqn:empDBFDindic} \end{equation}
From \cite{Gijbels15}'s analysis of \cite{Lopez09}'s band depth, though, one can understand how the non-continuous indicator function in this expression may prevent $\mu D(\chi,\widehat{P}_n)$ from being universally strongly consistent, even on small and well-behaved subsets $\Phi$. However, \cite{Gijbels15} showed how substituting $\indic{\cdot}$ with some smoothed version of it, fixes issues. Specifically, define $w: [0,\infty) \to [0,1]$ an {\it adjustment function}  such that $w$ is non-increasing on $[0,\infty)$, $w(0)=1$ and $\lim_{z \to \infty} w(z) = 0$ \citep[Definition 3]{Gijbels15}. Now replace $\indic{\cdot}$ in (\ref{eqn:empDBFDindic}) by $w\left(d(\chi,B_d(\chi_i,d(\chi_i,\chi_j)) \cap B_d(\chi_j,d(\chi_i,\chi_j)))\right)$, where the distance between $\chi$ and a subset $A$ of $\Ms$ is defined as $\inf_{\xi \in A} d(\chi,\xi)$. The choice $w(z) = \indic{z=0}$ reduces down to the original expression, while a continuous $w$ produces a smoothed version of it. \cite{Gijbels15} showed that indeed, for continuous $w$ as above, the `adjusted' version of the band depth is universally strongly consistent on every equicontinuous  subset  $ \Phi \subset \Ms$. We note that this result was derived in \cite{Gijbels15} for the case of the band depth on the space of continuous functions with supremum norm $(\Cs,\|\cdot\|_\infty)$ only -- in particular, its proof involves the Arzelà-Ascoli Theorem, a result specific to $(\Cs,\|\cdot\|_\infty)$. Although an attempt to extent this result to (\ref{eqn:empDBFDindic}) could be pursued,
\cite{Gijbels15} admitted that their adjustment is primarily motivated by theoretical considerations, but plays very little role in practice. Therefore, we will not consider this any further here. $\blacksquare$
\end{remark}

\ppn Finally, the obvious $U$-statistics structure of (\ref{eqn:empDBFD}) allows us to easily deduce, through an appropriate Central Limit Theorem, the asymptotic normality of $\mu D(\chi,\widehat{P}_n)$, a result that could be used for inference, for instance to build a confidence region for the `true' median element, i.e.\ the deepest element with respect to the population distribution $P$ \citep{Serfling15}.

\section{Data examples} \label{sec:realdat}

In this section, we illustrate the usefulness of the proposed metric depth $\mu D$ on 5 real data sets: two one-dimensional functional datasets (Sections \ref{subsec:canada} and \ref{subsec:realdatlips}), a bidimensional functional dataset (Section \ref{subsec:handwritting}), a symbolic data set (Section \ref{subsec:euro}) and a non-numeric data set (text) (Section \ref{sec:text}).


\subsection{Canadian weather data} \label{subsec:canada}

\begin{figure}[h]
	\centering
	\includegraphics[width=0.5\textwidth]{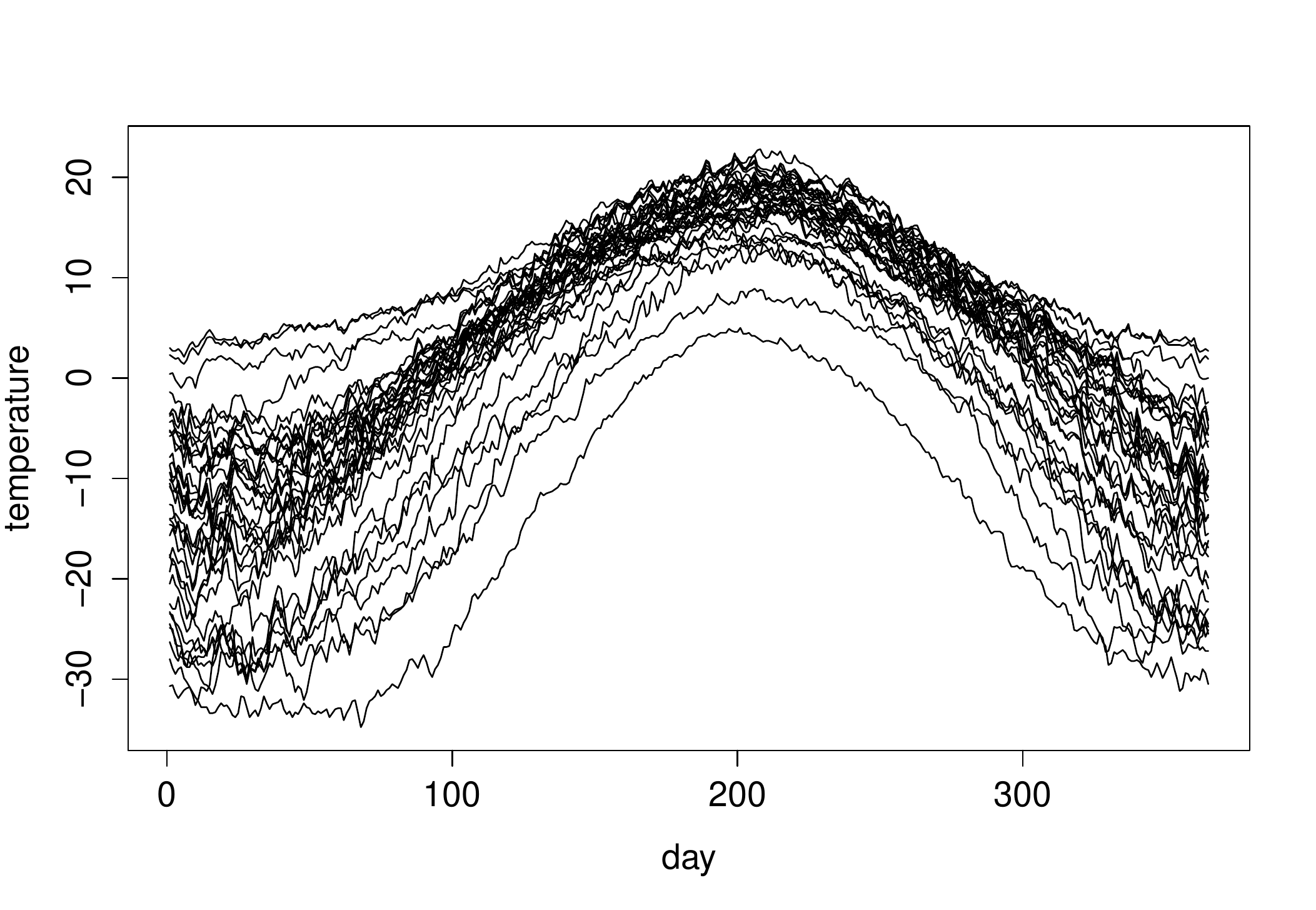}
\caption{Canadian weather data -- average daily temperatures at 35 stations.}
\label{fig:CWraw}
\end{figure}

\begin{figure}[h]
\centering
\includegraphics[width=.8\textwidth]{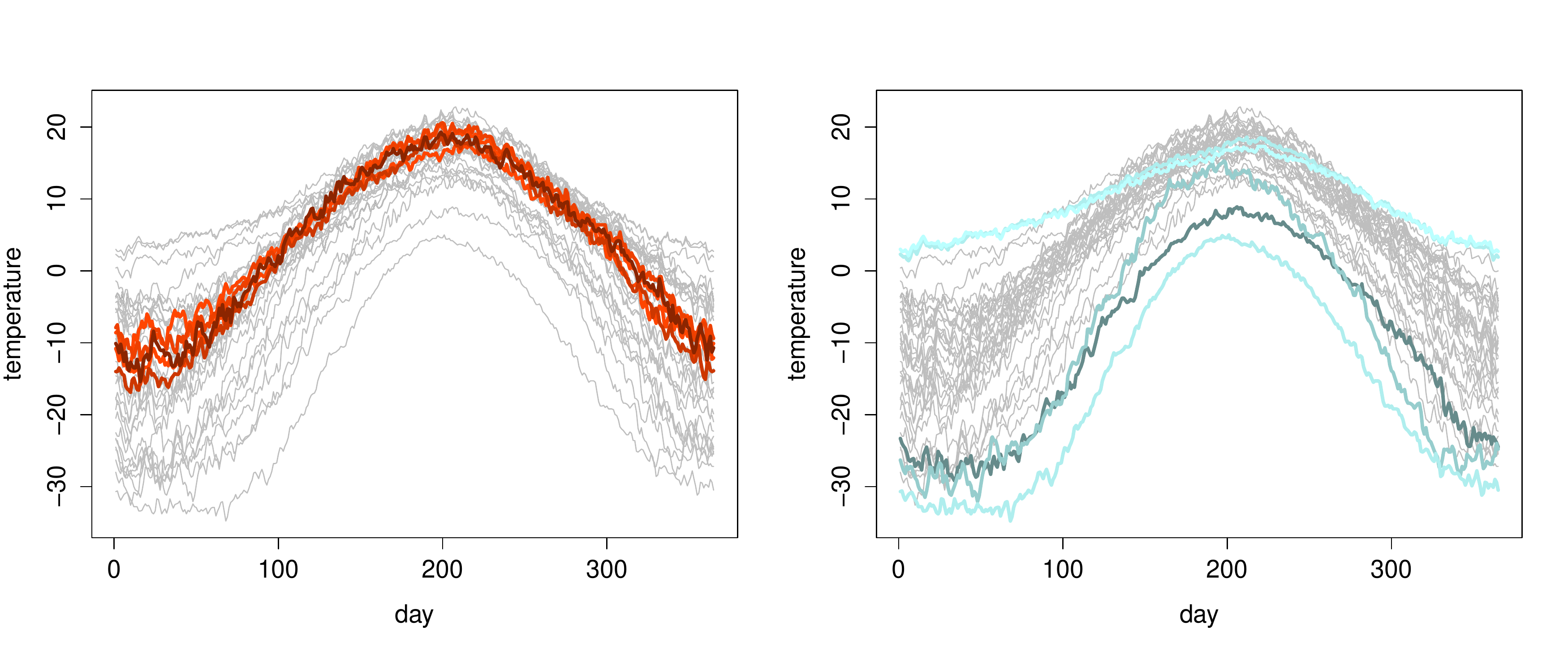}
\caption{Five deepest curves (left; the darkest curves are the deepest) and five least deep curves (right; the lightest curves are the least deep) according to (\ref{eqn:empDBFD}) with $d(\chi,\xi) = \|\chi-\xi\|_2$, the $L_2$ distance.}
\label{fig:CWdepth2}
\end{figure}

The Canadian temperature data set is a classical functional data set available from the {\sc R} package {\tt fda}. The data give the daily temperature records of 35 Canadian weather stations over a year (365 days, day 1 is 1st of January) averaged over 1960 to 1994, see Figure \ref{fig:CWraw}. First, the depth of the 35 curves with respect to the sample has been computed from the empirical functional metric depth (\ref{eqn:empDBFD}) with $d$ being the usual $L_2$ distance between two square-integrable functions, i.e.\ $d^2(\chi,\xi) = \int\left(\chi(t)-\xi(t) \right)^2\,dt$. The 5 deepest and least deep curves are shown in Figure \ref{fig:CWdepth2}. The suggested depth measure identifies the Sherbrooke (deepest curve), Thunder Bay, Fredericton, Quebec and Calgary stations as the most representative of a median Canadian weather, in terms of temperature. On the other hand, the most outlying curves are seen to be Resolute (least deep curve), Victoria, Vancouver, Inuvik and Iqaluit. It is visually obvious that those curves are much different to the others: Resolute, Inuvik and Iqaluit are Arctic stations, with much colder temperatures across the year than the other stations, while Vancouver and Victoria lie on the south Pacific coast of Canada and enjoy much milder winters. We can appreciate that Vancouver and Victoria are `shape outliers', whereas the Arctic stations are `location outliers'. These are equally easily flagged by the metric depth $\mu D$ -- this has to be stressed, as some functional depths have been shown to be able to identify one type of outlier but not the other, or vice-versa \citep{Serfling15}. 

\ppn Of course, the daily average temperature curves are particularly noisy, which could heavily affect the $L_2$-distances computed between pairs of curves, hence the whole calculation of the depths. One can deal with the roughness of those curves in different manners: first, one could use smoothed versions of the initial curves, for instance the monthly average temperatures as in \cite{Serfling15}; second, one could use for $d$ a distance less affected by such noise than the $L_2$ one, for instance the supremum ($L_\infty$) distance; finally one can expand the different curves in a certain basis and focus only on the first terms when assessing the proximity between them. We achieved that by expanding each curve in the empirical Principal Components basis \citep{Hall11} and keeping only the first two principal scores: the curves re-constructed from those two components only are indeed smooth approximations to the initial, rough curves. So, each curve is now represented by a point in the 2-dimensional space of the first two Principal Components, and the proximity between two curves quantified by the $L_2$-distance between the corresponding two points. In effect, this defines a pseudo-distance between the initial curves, see \citet[Section 3.4.1]{Ferraty06}. The depths assigned to each station according to these 4 methods are shown in Table \ref{tab:CW}. 
The four depth measures are in very good agreement, essentially identifying the same central and outlying curves. This shows that the depth measure $\mu D$ (\ref{eqn:FDBD}) and its empirical version (\ref{eqn:empDBFD}) are quite robust to any reasonable choice of $d$.


\begin{small}
\begin{table}[htb]
\centering
\begin{tabular}{|r| l l l l|r| l l l l|}
\hline
Station & $\mu D_2$ & $\mu D_\infty$ & $\mu D_2^\text{m}$ & $\mu D_\text{PCA}$ &
Station & $\mu D_2$ & $\mu D_\infty$ & $\mu D_2^\text{m}$ & $\mu D_\text{PCA}$\\
\hline
Sherbrooke&.514&.474&.516&.506 &
Thunder B.&.511&.513&.504&.496 \\
Fredericton&.504&.479&.504&.491 &
Quebec&.504&.513&.506&.497 \\
Calgary&.492&.375&.492&.503 &
Bagottville&.482&.457&.484&.476 \\
Edmonton&.474&.459&.472&.494 &
Arvida&.469&.476&.472&.476 \\
Regina&.429&.435&.427&.420 &
Charlottvl&.425&.435&.424&.435 \\
Pr.George&.422&.380&.425&.469 &
Ottawa&.410&.479&.408&.403 \\
Winnipeg&.403&.418&.405&.398 &
Pr.Albert&.400&.383&.400&.390 \\
Montreal&.383&.457&.380&.375 &
Halifax&.380&.348&.378&.368 \\
Whitehorse&.371&.378&.373&.395 &
The Pas&.360&.358&.358&.348 \\
Sydney&.324&.284&.333&.348 &
Uranium C.&.316&.257&.319&.311 \\
Toronto&.313&.363&.309&.309 &
Scheffervll&.294&.318&.292&.291 \\
St.Johns&.271&.237&.272&.257 &
London&.267&.324&.267&.261 \\
Yellowknife&.227&.207&.224&.200 &
Yarmouth&.208&.195&.208&.205 \\
Dawson&.207&.138&.208&.242 &
Churchill&.160&.242&.160&.160 \\
Kamloops&.150&.202&.148&.148 &
Pr.Rupert&.106&.108&.106&.109 \\
Iqaluit&.096&.134&.096&.096 &
Inuvik&.066&.064&.066&.066 \\
Vancouver&.045&.050&.045&.040 &
Victoria&.017&.002&.017&.018 \\
Resolute&0&0&0&0 &&&&& \\
\hline
\end{tabular}
\caption{Canadian weather data - metric depth measures for 4 different (pseudo-)distances $d$: $\mu D_2$: $L_2$ distance; $\mu D_\infty$: Supremum ($L_\infty$) distance; $\mu D_2^\text{m}$: $L_2$ distance on the average monthly temperature curves; $\mu D_\text{PCA}$: $L_2$ distance in the plane of the first two principal components.} 
\label{tab:CW}

\end{table}
\end{small}

\subsection{Lip movement data} \label{subsec:realdatlips}

\begin{figure}[h]
	\centering
	\includegraphics[width=0.5\textwidth]{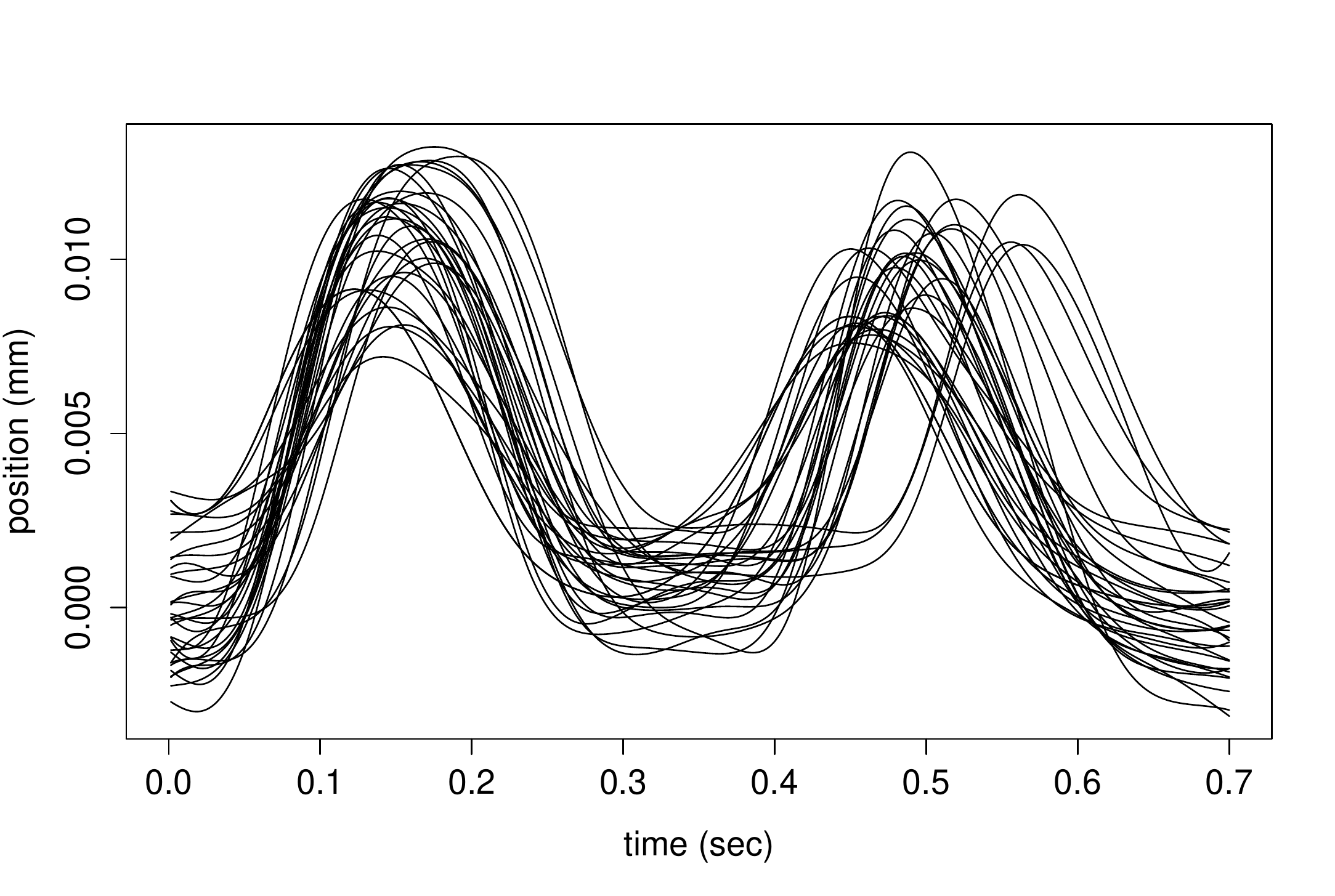}
\caption{Lip movement data.}
\label{fig:LipPos}
\end{figure}

\begin{figure}[htbp]
\begin{center}
\includegraphics[width=0.8\textwidth,height=4cm]{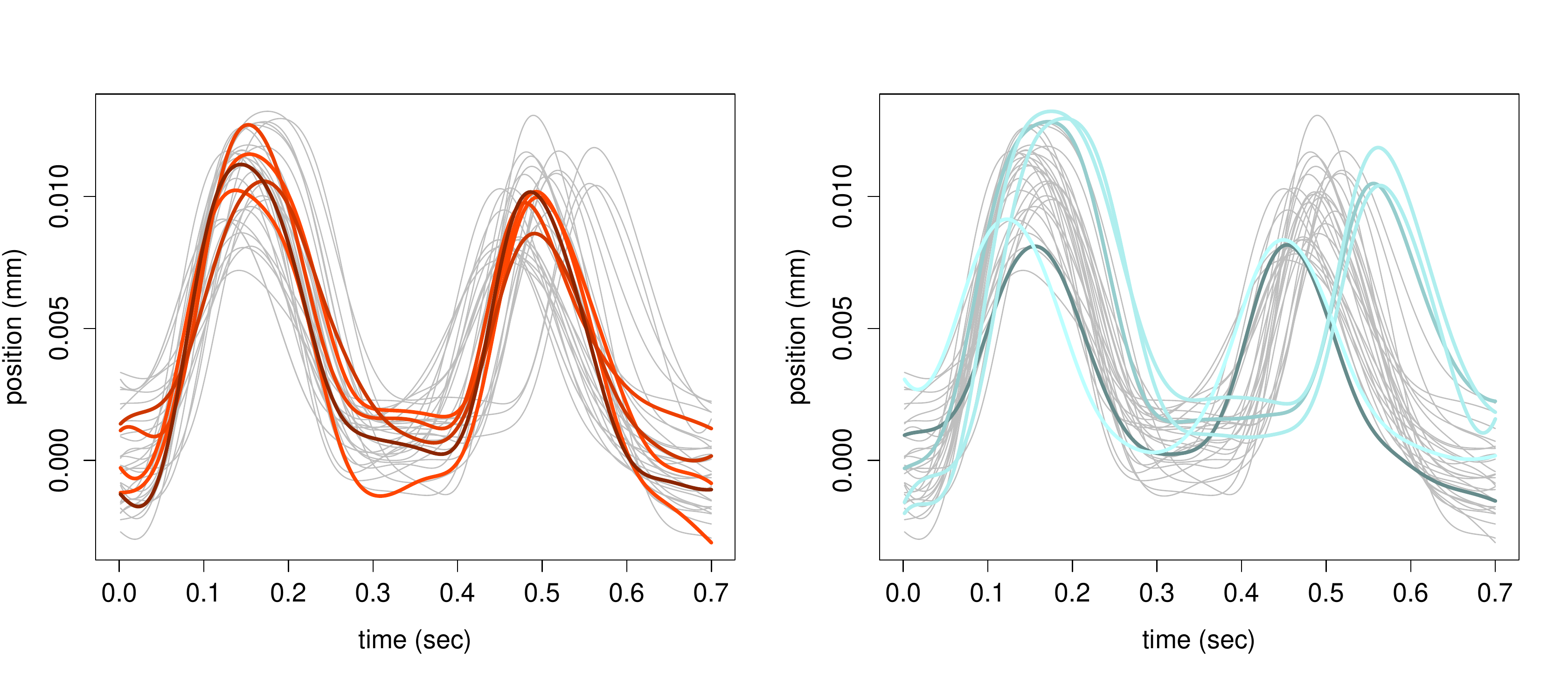}
\includegraphics[width=0.8\textwidth,height=4cm]{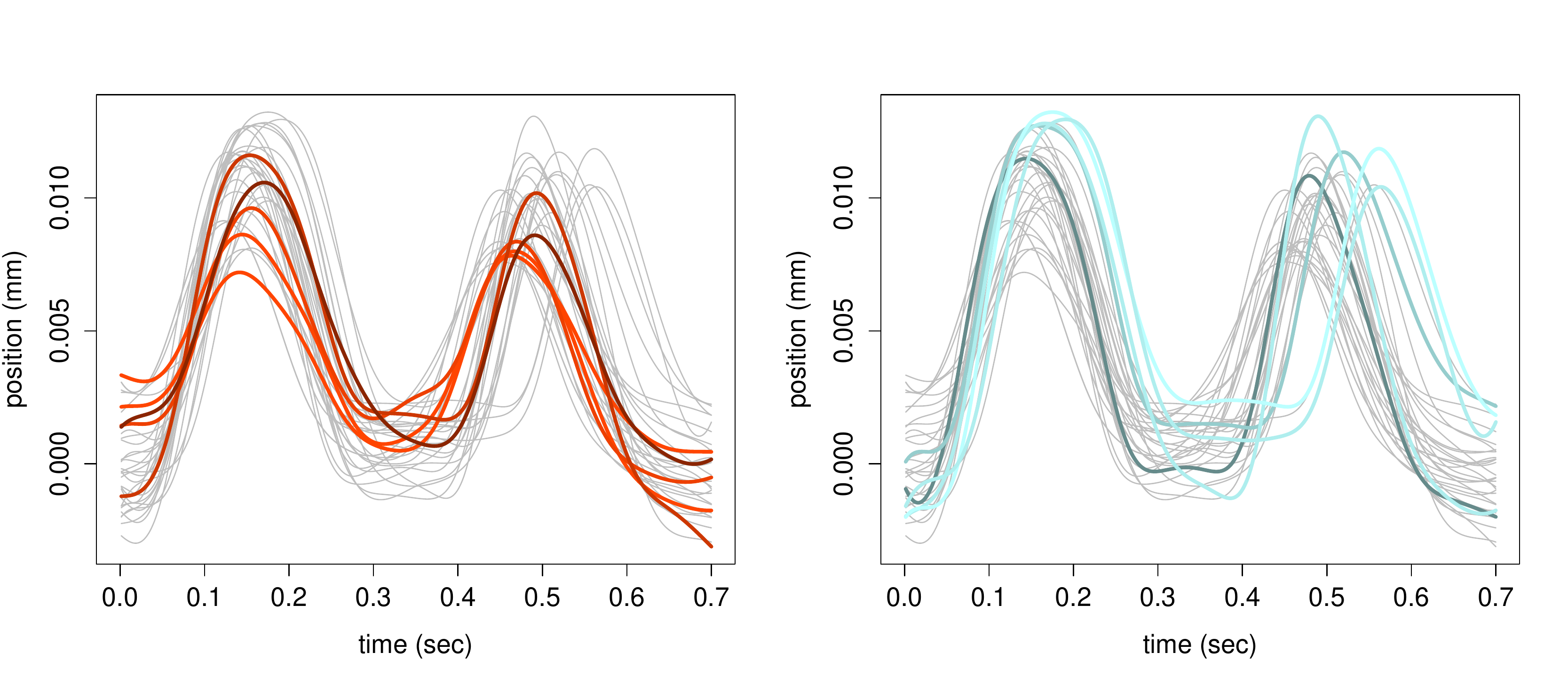}
\includegraphics[width=0.8\textwidth,height=4cm]{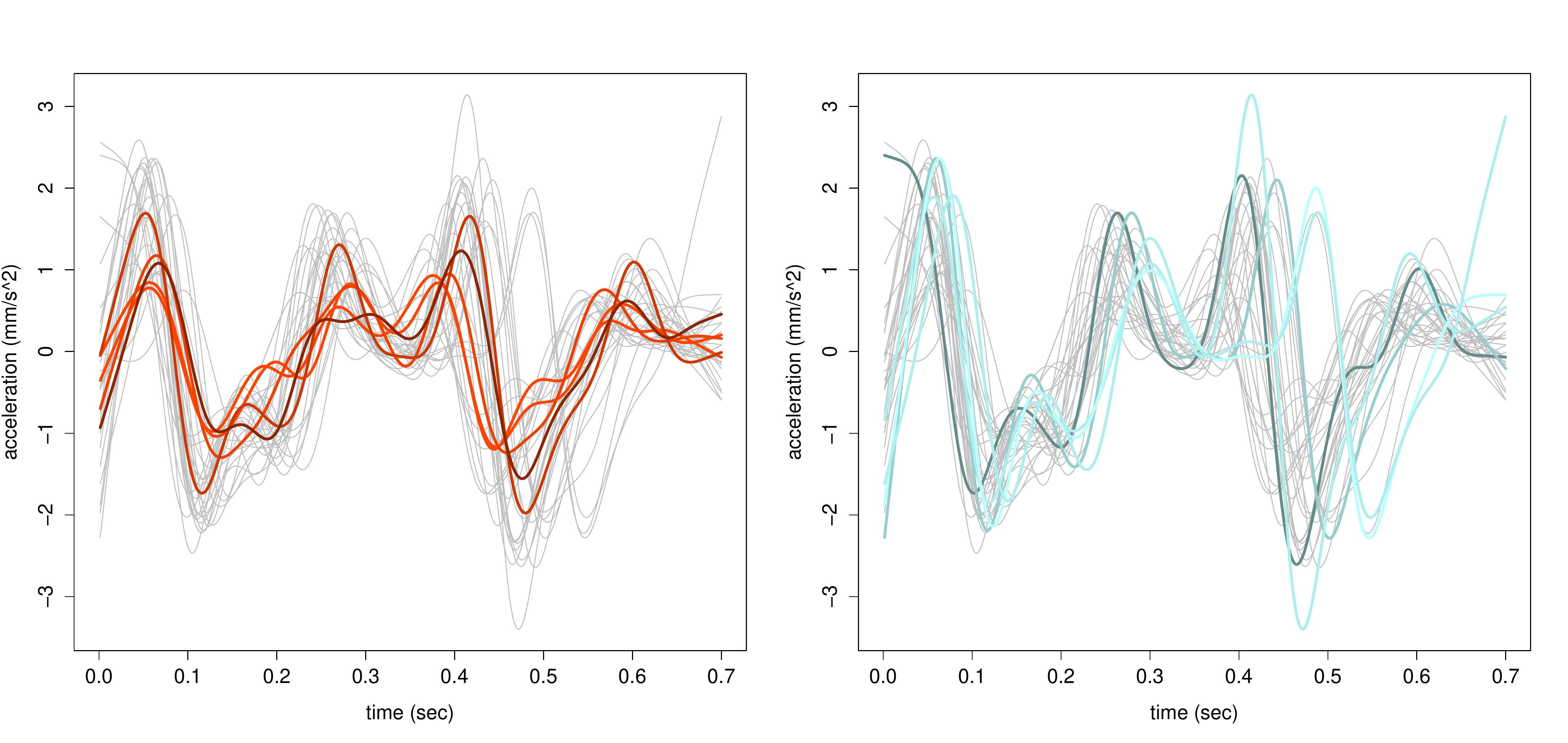}
\end{center}
\caption{Top and middle row:  Five deepest curves (left; the darkest curves are the deepest) and five least deep curves (right; the lightest curves are the least deep) according to (\ref{eqn:empDBFD}) with (i) $d(\chi,\xi) = \|\chi-\xi\|_2$, the $L_2$ distance between the curves (top row) and (ii) $d(\chi,\xi) = \|\chi''-\xi''\|_2$, the $L_2$ distance between the second derivatives of the curves (middle row). Bottom row: Five deepest acceleration curves (left; the darkest curves are the deepest) and five least deep acceleration curves (right; the lightest curves are the least deep).}
\label{fig:LP2}
\end{figure}

\cite{Malfait03} studied the relationship between lip movement and time of activation of different face muscles, see also \citet[Chapter 10]{Ramsay02} and \cite{Gervini08}. The study involved a subject saying the word `bob' 32 times and the movement of their lower lip was recorded each time. Those trajectories are shown in Figure \ref{fig:LipPos}, and all share the same pattern: a first peak corresponding to the firt /b/, then a plateau corresponding to the /o/ and finally a second peak for the second /b/. These functions being very smooth (actually, they are smoothed versions of raw data not publicly available), it seems natural to use again the classical $L_2$ distance for assessing their relative proximity. Hence, the respective depth of each curve with respect to the sample was obtained by (\ref{eqn:empDBFD}) with $d^2(\chi,\xi) = \int\left(\chi(t)-\xi(t) \right)^2\,dt$. The 5 deepest and 5 least deep curves are shown in the top row of Figure \ref{fig:LP2}. In particular, this depth identifies as outliers the three curves showing a second peak at a much later time than for the rest of the curves, which were already hived off by \cite{Gervini08}. The remaining two outlying curves show two peaks of lower amplitude than the others, with a second peak occurring earlier than the bunch.


\ppn Now, \cite{Malfait03}, in their original study, were more interested in the acceleration of the lip during the process rather than on the lip motion itself. The study aimed at explaining time of activation of face muscles, and the acceleration reflects the force applied to tissue by muscle contraction. Hence, in this application, it may be worth contrasting the lip trajectories in terms of their corresponding accelerations, that is, comparing the second derivatives of the position curves. The $L_2$ distance between the second derivatives of the curves is naturally a pseudo-distance between the initial curves \citep[Section 3.4.3]{Ferraty06}, which can be used in (\ref{eqn:empDBFD}). The 5 deepest and 5 least deep curves, according to (\ref{eqn:empDBFD}) based on the `acceleration' pseudo-distance, are shown in the middle row of Figure \ref{fig:LP2} and differ from those in the first row of Figure \ref{fig:LP2}. Naturally, the focus here is no more on the exact position of the curves, but rather on the more fundamental underlying dynamics. For instance, the 5 deepest curves show a first peak of distinctly different heights, but in terms of their second derivatives, they are in fact quite similar and representative of the sample (bottom row of Figure \ref{fig:LP2}), and that is what matters in \cite{Malfait03}'s study. As argued in Section \ref{sec:intro}, the flexibility of $\mu D$ (\ref{eqn:FDBD}) in terms of the choice of $d$ allows the analyst to tailor the depth measure to the given factors and the goal of the analysis.

\subsection{Handwriting data} \label{subsec:handwritting}

\begin{figure}[h]
\centering
\includegraphics[width=0.5\textwidth]{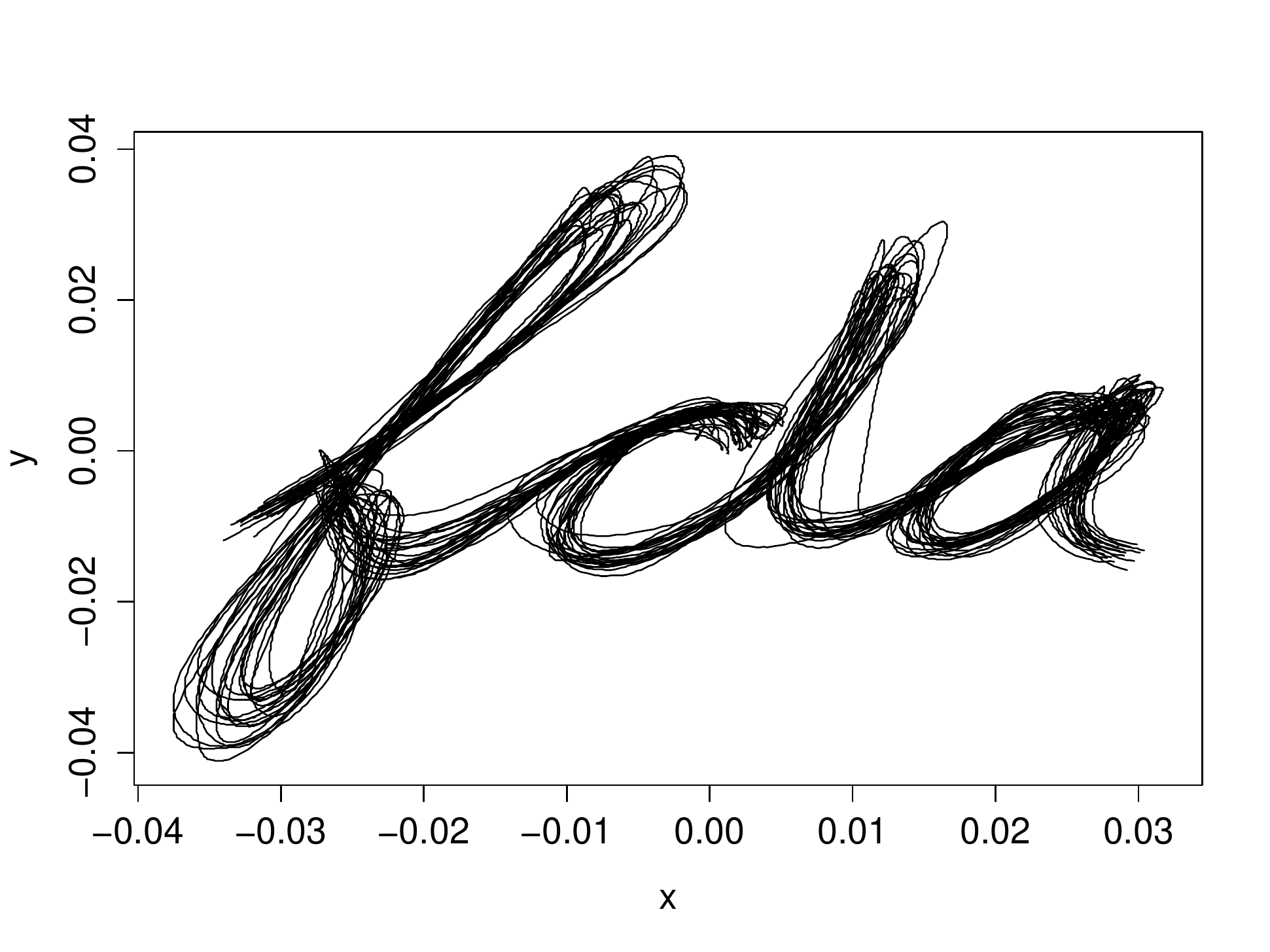}
\caption{ Handwriting data.}
\label{fig:fdaraw}
\end{figure}

The `handwriting' data set consists of twenty replications of the printing of the three letters `{\it fda}' by a single individual. The position of the tip of the pen has been sampled 200 times per second. The data, available in the {\sc R} package {\tt fda}, have already been pre-processed so that the printed characters are scaled and oriented appropriately, see Figure \ref{fig:fdaraw}.

\begin{figure}[h]
\centering
\includegraphics[width=0.6\textwidth]{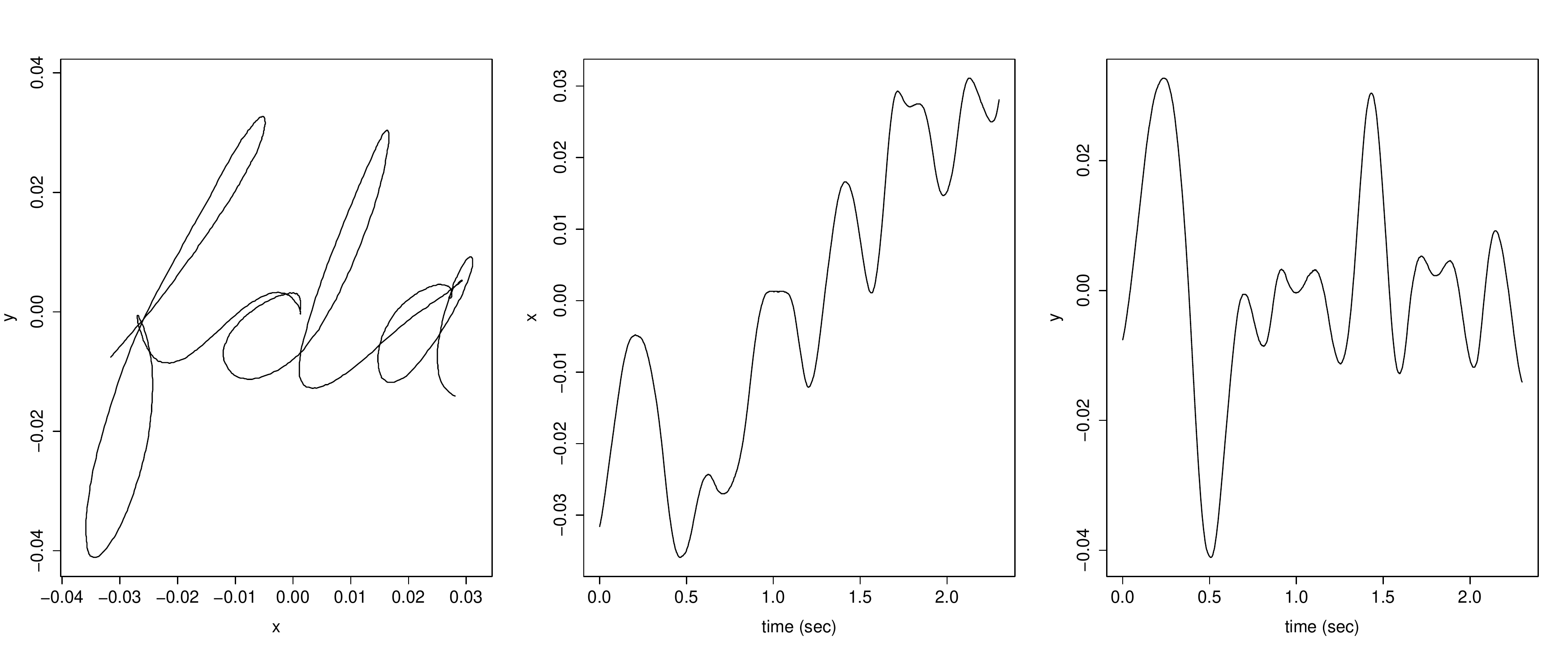}
\caption{One instance of the handwriting data, and its $x$- and $y$-components.}
\label{fig:fdaXY}
\end{figure}

\begin{figure}[h]
\centering
\includegraphics[width=0.7\textwidth]{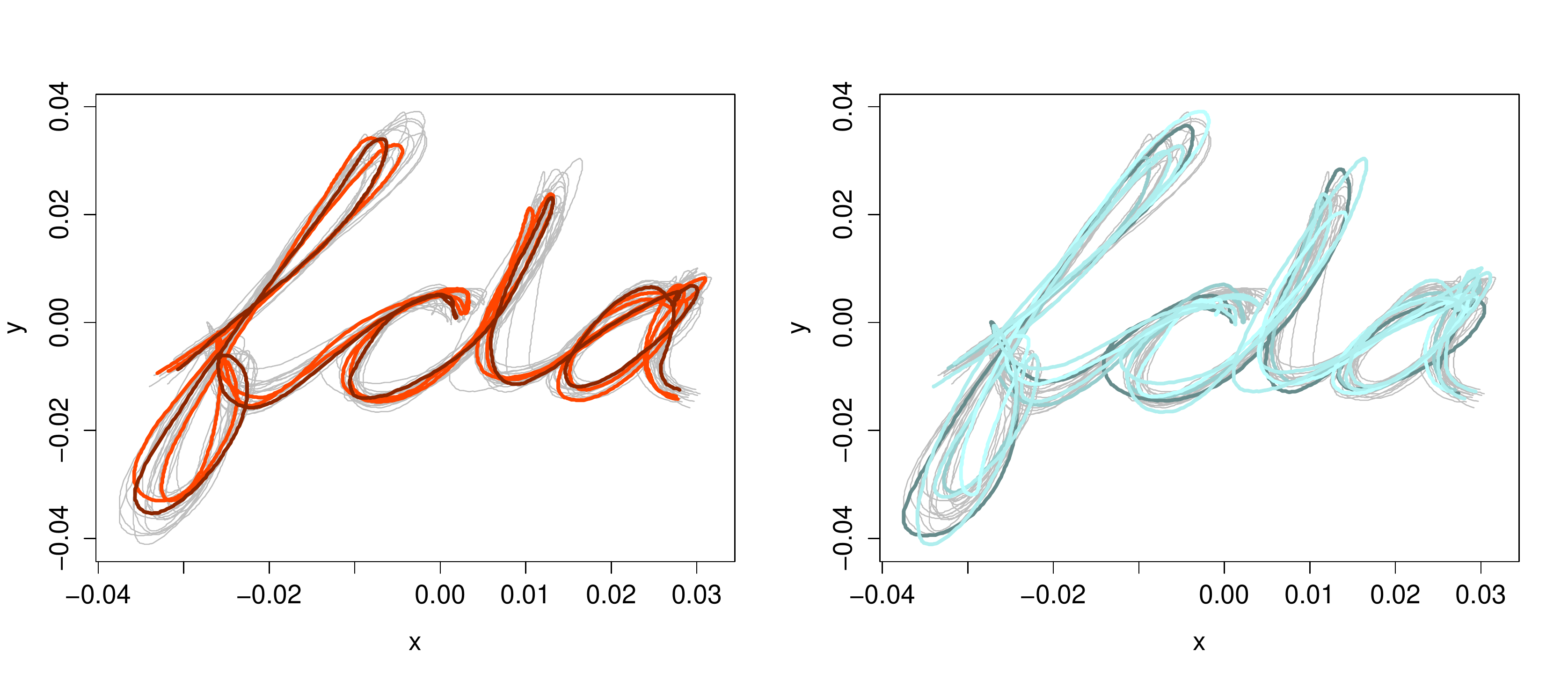}
\caption{Five deepest curves (left; the darkest curves are the deepest) and five least deep curves (right; the lightest curves are the least deep) according to (\ref{eqn:empDBFD}) with $d(\chi,\xi)$ being the $L_2$ distance (\ref{eqn:bivL2}) on $L_2(T) \times L_2(T)$.}
\label{fig:fdadeep}
\end{figure}

\ppn These data are essentially bivariate functional data. Indeed, each instance $\chi$ of the word `fda' arises through the simultaneous realisation of two components $(\chi_X(t),\chi_Y(t))$, where $\chi_X(t)$ and $\chi_Y(t)$ give the position along the horizontal axis and the  vertical axis, respectively, of the pen at time $t$. This is illustrated for one instance of `fda' in Figure \ref{fig:fdaXY}. Hence, an appropriate functional metric space here could be $(\Ms,d)$ with $\Ms = L_2(T) \times L_2(T)$, $T=[0,2.3]$ (the time interval on which the position of the pen was recorded) and $d$ being the Euclidean distance on $L_2(T) \times L_2(T)$ whose square is defined by
\begin{equation} d^2(\chi,\xi) = \|\chi_X - \xi_X \|_2^2 + \|\chi_Y - \xi_Y \|_2^2 = \int_T\left(\chi_X(t)-\xi_X(t) \right)^2\,dt + \int_T\left(\chi_Y(t)-\xi_Y(t) \right)^2\,dt.  \label{eqn:bivL2} \end{equation}
This distance can be used directly in (\ref{eqn:empDBFD}) to identify the 5 deepest and 5 least deep instances of `fda', see Figure \ref{fig:fdadeep}. The bivariate nature of the data at hand does not cause any particular complication and the definition (\ref{eqn:FDBD}) need not be re-adapted to this case. Again, the so-defined depth only focuses on the `drawings' {\it fda} themselves, and identifies the deepest instances. However, it was argued in the related literature that the tangential acceleration of the pen during the process was also a key element to analyse for understanding the writing dynamics, for instance for discriminating between genuine handwritings and forgeries \citep{Geenens11a,Geenens11b}. As in Subsection \ref{subsec:realdatlips}, one could therefore use (\ref{eqn:empDBFD}) with $d$ a pseudo-distance assessing the proximity between two instances of {\it fda} through their tangential acceleration curves only, if that was to be the focus of the analysis.

\subsection{Age distribution in European countries} \label{subsec:euro}

Symbolic Data Analysis (SDA) has recently grown as a popular research field in statistics \citep{Billard03,Billard07}. Indeed the intractably large `Big Data' sets often need to be summarised so that the resulting summary datasets are of a manageable size, and so-called `symbolic data' typically arise from such a process. No longer formatted as single values like classical data, they are meant to be `aggregated' variable typically represented by lists, intervals, histograms, distributions and the like. In this section we give a closer look at a `distribution-valued' symbolic data set. Specifically, we analyse the distribution of the age of the population of the 44 european countries (see Table \ref{tab:eurpop}).

\begin{figure}[h]
\centering
\includegraphics[width=0.5\textwidth]{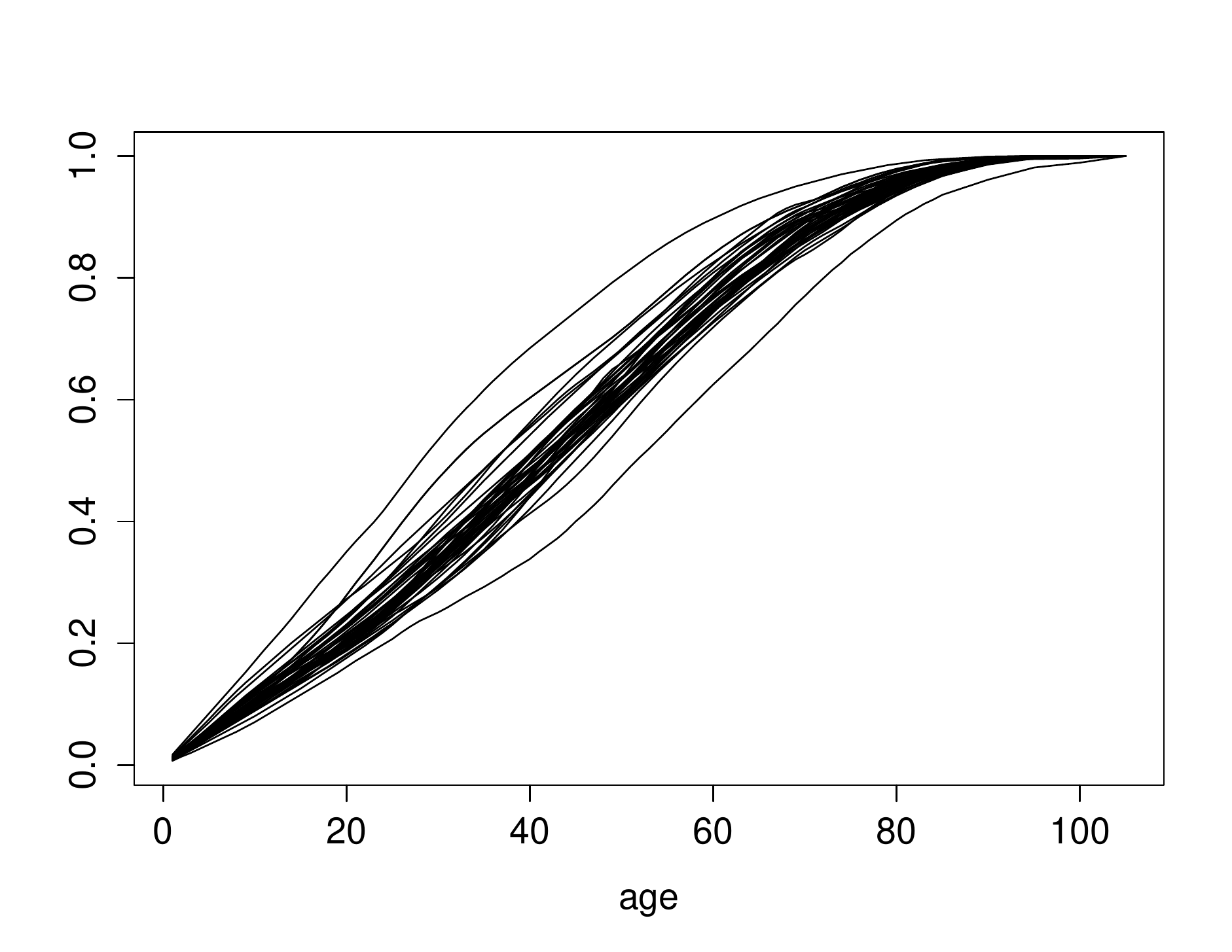}
\caption{Age distribution in European countries}
\label{fig:europe-pop}
\end{figure}

\ppn The 2017 data were obtained from the US Census bureau ({\tt www.census.gov/population/\\international/data/}). Typically, the population distribution for a given country is presented under the form of a population pyramid (that is, a histogram), from which a proper distribution function for population age can easily be extracted \citep{Kosmelj11}. Hence, each country (here: `individual', also called `concept' in the SDA literature) is characterised by a distribution. Figure \ref{fig:europe-pop} displays the sample of age distributions. 
Here we will use the suggested metric depth $\mu D$ to analyse which countries are most representative of the `European' age distribution, and which countries can be regarded as `outliers' in that respect.

\begin{small}
\begin{table}[H]
\centering
\begin{tabular}{|l|l|l|l|l|l|l|l|}
\hline
 Switzerland              &.59  &
 Liechtenstein            &.52  &
 Hungary                  &.519 &
 Malta                    &.508 \\
 Czech Republic           &.503 &
 Ukraine                  &.498 &
 Netherlands              &.488 &
 Croatia                  &.478 \\
 Portugal                 &.477 &
 Poland                   &.474 &
 Belgium                  &.451 &
 Serbia                   &.449 \\
 Denmark                  &.434 &
 Romania                  &.422 &
 UK		           &.421 &
 Belarus                  &.406 \\
 Spain                    &.406 &
 Estonia                  &.401 &
 Bulgaria                 &.36  &
 Montenegro               &.36  \\
 Slovakia                 &.349 &
 Latvia                   &.348 &
 Sweden                   &.322 &
 Lithuania                &.319 \\
 Luxembourg               &.318 &
 Russia                   &.292 &
 Norway                   &.28  &
 Austria                  &.277 \\
 Bosnia-Herz.   &.276 &
 France                   &.27  &
 Finland                  &.261 &
 San Marino               &.228 \\
 Andorra                  &.221 &
 Macedonia                &.201 &
 Slovenia                 &.178 &
 Moldova                  &.151 \\
 Greece                   &.141 &
 Iceland                  &.14  &
 Ireland                  &.088 &
 Italy                    &.087 \\
 Albania                  &.044 &
 Germany                  &.044 &
 Kosovo                   &0     &
 Monaco                   &0 \\
\hline
\end{tabular}
\caption{Age distribution in European countries -- metric depth for the age distributions of the 44 European countries, based on the Wasserstein distance.} 
\label{tab:eurpop}
\end{table}
\end{small}

\ppn The data being here distribution functions of nonnegative variables, $\Ms$ can be identified with a space of distribution functions supported on $\R^+$, i.e.\ a space of nondecreasing c\`adl\`ag functions $F$ with $F(0)=0$ and $\lim_{t \to \infty} F(t) =1 $, equipped with an appropriate distance. The Wasserstein distance has proved useful for a wide range of problems explicitly involving distribution functions \citep{Rachev84,Panaretos20}, hence seems a natural choice in this setting as well. For some $r \geq 1$, the Wasserstein distance between two distributions $F$ and $G$ whose $r$th moments exist, is defined as
\[d_r(F,G) = \inf_{(X,Y)\sim (F,G)} \{E\left( |X-Y|^r\right)\}^{1/r}, \]
where the infimum is taken over the set of all joint bivariate distributions whose marginal distributions are $F$ and $G$ respectively. Properties of this distance are described in \cite{Major78} and \cite{Bickel81}. In particular, it is known that $d_r(F,G)$ is essentially the usual $L_r$-distance between the quantile functions $F^{-1}$ and $G^{-1}$ over $[0,1]$. Also, it is known that convergence in the Wasserstein distance is equivalent to convergence in distribution together with convergence of the first $r$ moments. Hence, the distance $d_r$ quantifies the proximity between two distributions through both their general appearance and the values of their moments. In what follows, we take $r=2$, hence we consider functional data in $(\Ms_2,d_2)$, $\Ms_2$ being the space of all probability distribution functions with finite second moment.

\begin{figure}[h]
\centering
\includegraphics[width=0.8\textwidth]{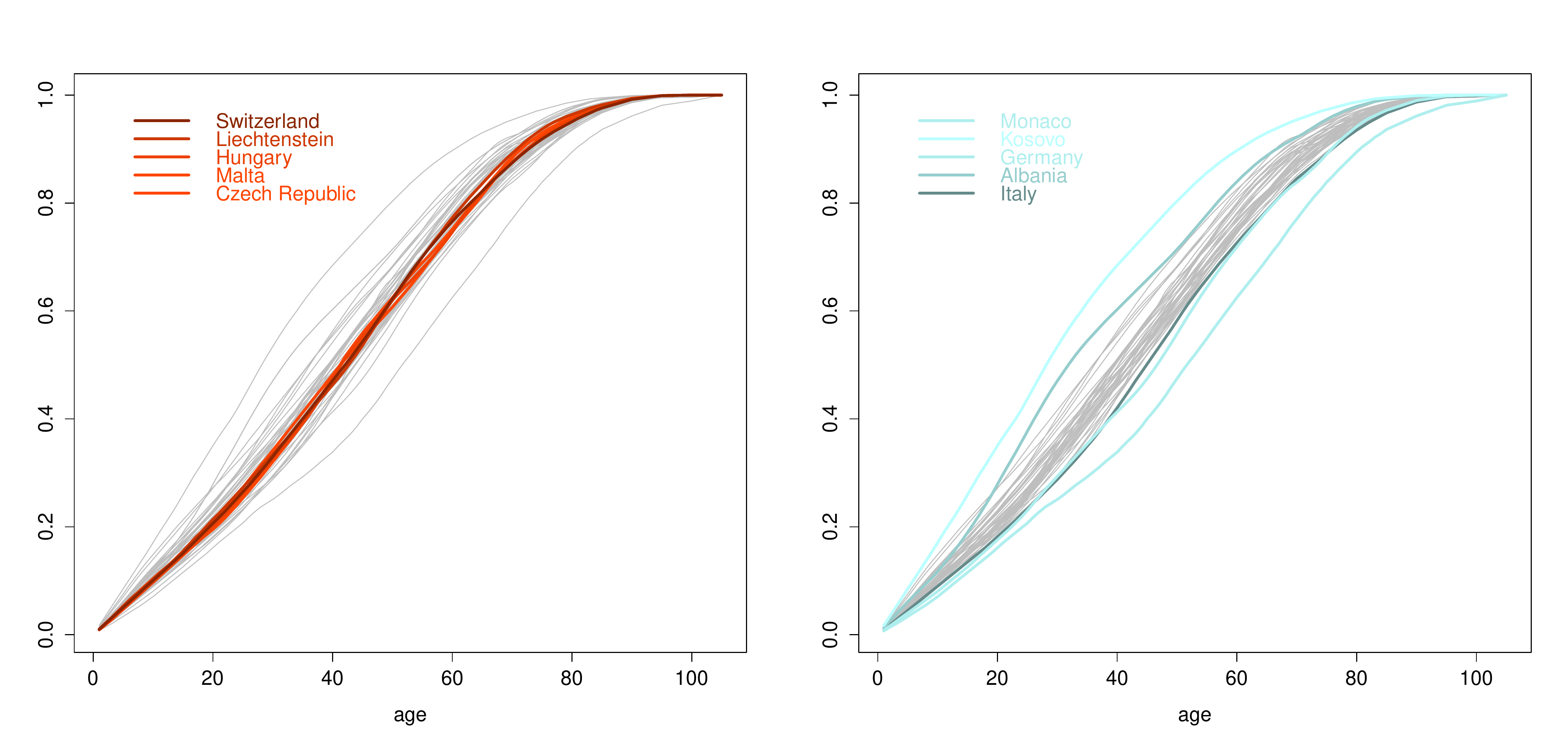}
\caption{Five deepest age distributions (left; the darkest curves are the deepest) and five least deep age distributions (right; the lightest curves are the least deep) according to (\ref{eqn:empDBFD}) with $d$ being the Wasserstein distance $d_2$ between distributions.}
\label{fig:europe-pop-deep}
\end{figure}

\ppn The flexibility of (\ref{eqn:FDBD}) allows us to base $\mu D$ on the Wasserstein distance so as to define a depth measure specific to distribution functions without any difficulty. The `Wasserstein-depths' of the 44 countries are given in Table \ref{tab:eurpop}. 
The 5 deepest and least deep age distributions are shown in Figure \ref{fig:europe-pop-deep}. The deepest distribution, hence the most representative of the age distributions in Europe, appears to be that of Switzerland, a country located at the very heart of Europe, in-between the Western and Eastern countries, and in-between the Northern countries and the Southern countries, at the meeting point between the `Germanic' world (Germany, Austria) and the `Latin' world (France, Italy). From that perspective, Switzerland can be regarded as really representative of a `median' European country on many aspects. On the other hand, the Wasserstein-metric depth is null for Kosovo and Monaco, and indeed, the distributions for those two countries clearly lie outside the bunch of the other distributions. Monaco is a micro, mild-climate (and incidentally, tax haven) state which attracts a large amount of rich retirees from all over the continent (if not the world), hence its population is globally much older than for other countries and its age distribution is below the others. Monaco set aside, Germany and Italy show globally the oldest population of Europe. Kosovo was still recently at the heart of an armed conflict in the Balkans, which explains the low proportion of older people in that country and the position of its age distribution above all the others. To some extent, this also explains the outlyingness of Albania's curve. In any case, this example illustrates that one can readily define a depth measure tailored for distribution curves, which paves the way for developing rank-like procedures in Symbolic Data Analysis as well.

\subsection{Authorship attribution by intertextual distance} \label{sec:text}

Author identification on an unknown or doubtful text is one of the oldest statistical problems applied to literature. Here the capability of the proposed metric depth is illustrated within that framework. William Shakespeare and Thomas Middleton were contemporaries (late 16th-early 17th centuries), and their oeuvre are often compared. In that aim, \cite{Merriam03} examined 9 Middleton plays and 37 Shakespeare texts, and computed between each pair of them the so-called `inter-textual distance' proposed by \cite{Labbe01}.\footnote{It is not the purpose of this paper to describe how this index is computed or what it represents; neither do we imply that it is the panacea for the considered problem -- for that matter, it has been criticised \citep{Viprey06}. Here we use it in an illustrative purpose only.}  Although the entities of interest are here purely non-numerical (famous literary pieces), the obtained matrix of distances allows us to outline the relative position of each text -- and this is essentially all what is needed for $\mu D$ to come into play. 

\ppn As an example, Table \ref{tab:Midd} (recovered from Appendix 2 in \cite{Merriam03}) reports the `inter-textual' distances between the 9 essential plays of Middleton. Computing the empirical metric-depth (\ref{eqn:empDBFDindic}) on each of this entry in the `Middleton sample' reveals that the two deepest observations are `More Dissemblers Besides Women' and  `A Trick to Catch the Old One' (both get a depth of 0.4167). They may, therefore, be considered as the most typical Middleton plays (as long as the `inter-textual' distance is the relevant metric).

\begin{table}
\[ \begin{array}{c | c c c c c c c c c}
 & \text{\it Phn} & \text{\it Mad} & \text{\it Trk} & \text{\it Pur} & \text{\it Alm} & \text{\it CMC} & \text{\it Dis} & \text{\it Val} & \text{\it WBW} \\
 \hline 
\text{\it Phn} &0.000 &0.312 &0.301 &0.315 &0.335 &0.343 &0.319 &0.330 &0.322 \\
\text{\it Mad}	 &0.312 &0.000 &0.314 &0.344 &0.319 &0.328 &0.326 &0.339 &0.325 \\
\text{\it Trk}	 &0.301 &0.314 &0.000 &0.310 &0.314 &0.326 &0.335 &0.338 &0.330 \\
\text{\it Pur}	 &0.315 &0.344 &0.310 &0.000 &0.338 &0.340 &0.337 &0.346 &0.338 \\
\text{\it Alm}	 &0.335 &0.319 &0.314 &0.338 &0.000 &0.314 &0.313 &0.333 &0.344 \\
\text{\it CMC}	 &0.343 &0.328 &0.326 &0.340 &0.314 &0.000 &0.339 &0.349 &0.311 \\
\text{\it Dis}	 &0.319 &0.326 &0.335 &0.337 &0.313 &0.339 &0.000 &0.318 &0.284 \\
\text{\it Val}	 &0.330 &0.339 &0.338 &0.346 &0.333 &0.349 &0.318 &0.000 &0.320 \\
\text{\it WBW}	 &0.322 &0.325 &0.330 &0.338 &0.344 &0.311 &0.284 &0.320 &0.000 
\end{array} \]
	\caption{Matrix of `inter-textual' distances between 9 essential plays by Thomas Middleton: {\it Phn}: `The Phoenix'; {\it Mad}: `A Mad World, My Masters'; {\it Trk}: `A Trick to Catch the Old One'; {\it Pur}: `The Puritan'; {\it Alm}: `The Almanac'; {\it CMC}: `A Chaste Maid in Cheapside'; {\it Dis}: `More Dissemblers Besides Women'; {\it Val}: `The Nice Valour'; {\it WBW}: `Women Beware Women'.} \label{tab:Midd}
\end{table}

\ppn This time focusing on the 37 Shakespeare texts only, `Antony and Cleopatra' is identified as Shakespeare's most typical text; i.e., the deepest among the considered sample (depth: 0.5255) -- see Table \ref{tab:Shak-depth} (left column). The following most representative of Shakespeare plays are `The Tempest' (0.5135), `Othello' (0.5030) and `Romeo and Juliet' (0.5015). The most outlying piece of work is the verset part of `Henry V' (depth: 0), which tends to confirm a common conjecture hold by many experts on Shakespeare's oeuvre: the verset part of `Henry V' was not written by Shakespeare himself, but by Christopher Marlowe \citep{Merriam02}. 

\ppn Now, if we computed the metric depth of the 9 Middleton's plays in Shakespeare's sample, all would receive depth 0 -- all are `outlying' in Shakespeare's oeuvre. This clearly indicates that Middleton's work cannot be confused with Shakespeare's, and it should be easy to assign a new piece of text to one or the other based on $\mu D$. Further, it is interesting to analyse the depth of each text in a combined sample made up both the works of Middleton and Shakespeare. In particular, some of Shakespeare's texts which have a low depth in the `Shakespeare's only' sample, see their depth increase by large in the combined sample. This indicates that these pieces may have a strong Middleton flavour, to some extent. This hypothesis is confirmed for at least one of those plays: `Timon of Athens' sees its depth increase from 0.1141 to 0.3971 if one includes Middleton's works in the reference sample; and indeed, extensive research on the topic has provided ample evidence that Middleton wrote approximately one third of that play \citep{Taylor87}. 

\ppn Note that computing and comparing the depth of certain observations in two different samples is the spirit of the $DD$-plot and the $DD$-classifier proposed by \cite{Li12}. These procedures can naturally be used in conjunction with the metric depth $\mu D$, enabling similar powerful depth-based analyses in abstract metric spaces.

\begin{table}
	\centering
	\begin{tabular}{l | c | c}
		&   $\mu D$ - Shakespeare &  $\mu D$ - combined \\
		\hline
		The Two Gentlemen of Verona &0.3784 &0.5072 \\
		The Taming of the Shrew &0.2192 &0.4396 \\
		Henry VI - Part II &0.2508 &0.2097 \\
		Henry VI - Part III &0.0526 &0.0425\\
		Henry VI - Part I &0.1532 &0.1227\\
		Titus Andronicus &0.1937 &0.1594\\
		Richard III &0.4760 &0.4570\\
		The Comedy of Errors &0.4595 &0.5353\\
		Love's Labour's Lost &0.4910 &0.4889\\
		A Midsummer Night's Dream &0.3859 &0.3691\\
		Romeo and Juliet  &0.5015 &0.5227\\
		Richard II &0.1006 &0.0841\\
		King John &0.2267 &0.2048\\
		The Merchant of Venice &0.4189 &0.5179\\
		Henry IV, Part I &0.4129 &0.4058\\
		The Merry Wives of Windsor &0.3829 &0.4483\\
		Henry IV, Part II &0.3619 &0.4744\\
		Much Ado about Nothing &0.2252 &0.4444\\
		Henry V (prose part) &0.3078 &0.2870\\
		Henry V (verset part) &0.0000 &0.0000\\
		Julius Caesar &0.4294 &0.3990\\
		As You Like It &0.2763 &0.4792\\
		Hamlet &0.4715 &0.4473\\
		Twelfth Night &0.0075 &0.3488\\
		Troilus and Cressida &0.3333 &0.2850\\
		Measure for Measure &0.3378 &0.5063\\
		Othello &0.5030 &0.5585\\
		All's Well that Ends Well &0.0450 &0.3710\\
		Timon of Athens &0.1141 &0.3971\\
		King Lear &0.4640 &0.5478\\
		Macbeth &0.3649 &0.3121\\
		Antony and Cleopatra &0.5255 &0.5401\\
		Coriolanus &0.4655 &0.4309\\
		The Winter's Tale &0.2462 &0.4261\\
		Cymbeline &0.1607 &0.4184\\
		The Tempest &0.5135 &0.5246\\
		Henry VIII &0.4099 &0.5024\\
		\hline
	\end{tabular}
	\caption{37 Shakespeare's plays (shown in chronological order) -- empirical $\mu D$ in the sample of Shakespeare's plays only (left column) and in the sample of combined Shakespeare's and Middleton's works (right column).}
	\label{tab:Shak-depth}	
\end{table}	

\section{Conclusion} \label{sec:ccl}

In this paper, we have proposed a new statistical depth function, called `metric depth' or just $\mu D$, defined in an abstract metric space. It is explicitly constructed on a certain distance $d$ that must be chosen by the analyst, which allows them to tailor the depth to the data at hand and to the ultimate goal of the analysis. This offers an unmatched flexibility about the range of problems and applications that can be addressed using the said depth measure. The usefulness of $\mu D$ has been illustrated on several real data sets, including one in the emergent field of Symbolic Data Analysis and an application in text mining (authorship attribution). 
Rejuvenating an old idea of \cite{Bartoszynski97}, its definition is very intuitive: the depth of a functional point $\chi$ with respect to a distribution $P$ is the probability to find it `between' two functional objects $\Xs_1$ and $\Xs_2$ randomly generated from $P$, `between' meaning here that $\chi$ belongs to the intersection of the two open $d$-balls $B_d(\Xs_1,d(\Xs_1,\Xs_2))$ and $B_d(\Xs_2,d(\Xs_1,\Xs_2))$. This definition is natural and 
enjoys many pleasant properties. 

\section*{Acknowledgements} 

Gery Geenens' research was supported by a Faculty Research Grant from the Faculty of Science, UNSW Sydney, Australia. Alicia Nieto-Reyes' research was funded by  the Spanish Ministerio de Ciencia, Innovaci\'on y Universidades grant number MTM2017-86061-C2-2-P.

\bibliographystyle{Chicago}

\newpage

\appendix

\section{Appendix}

Here we give three counter-examples for illustrating that \cite{Liu11}'s Euclidean `lens depth' does not satisfy \cite{Zhuo00}'s properties P2 `Maximality at centre'  and P3 `Monotonicity relative to deepest point' for centrally symmetric distributions -- indeed, these two properties are related. For simplicity, we work in $\R^2$.

\begin{figure}[hbt]
	\includegraphics[width=0.2425\linewidth]{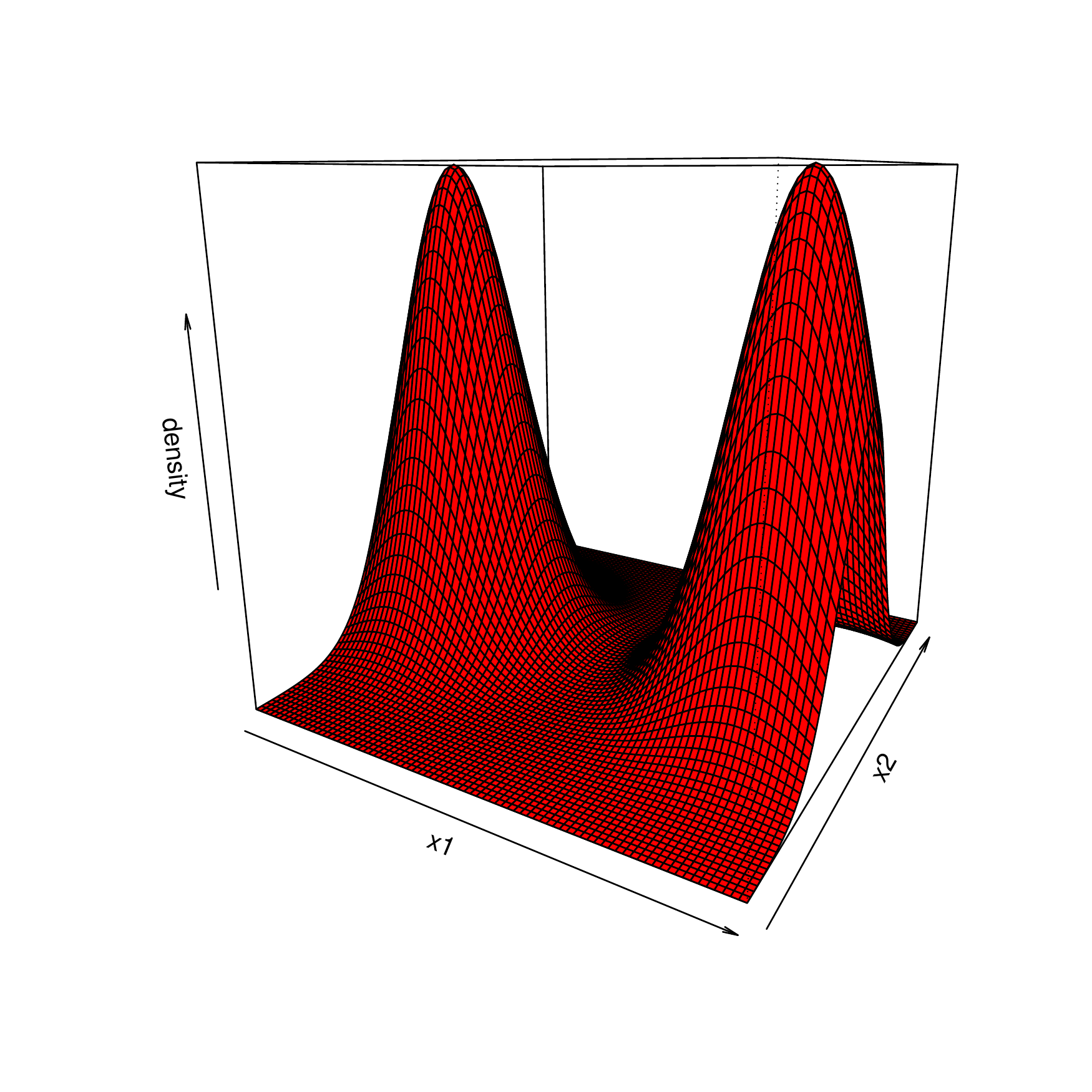}
	\includegraphics[width=0.2425\linewidth]{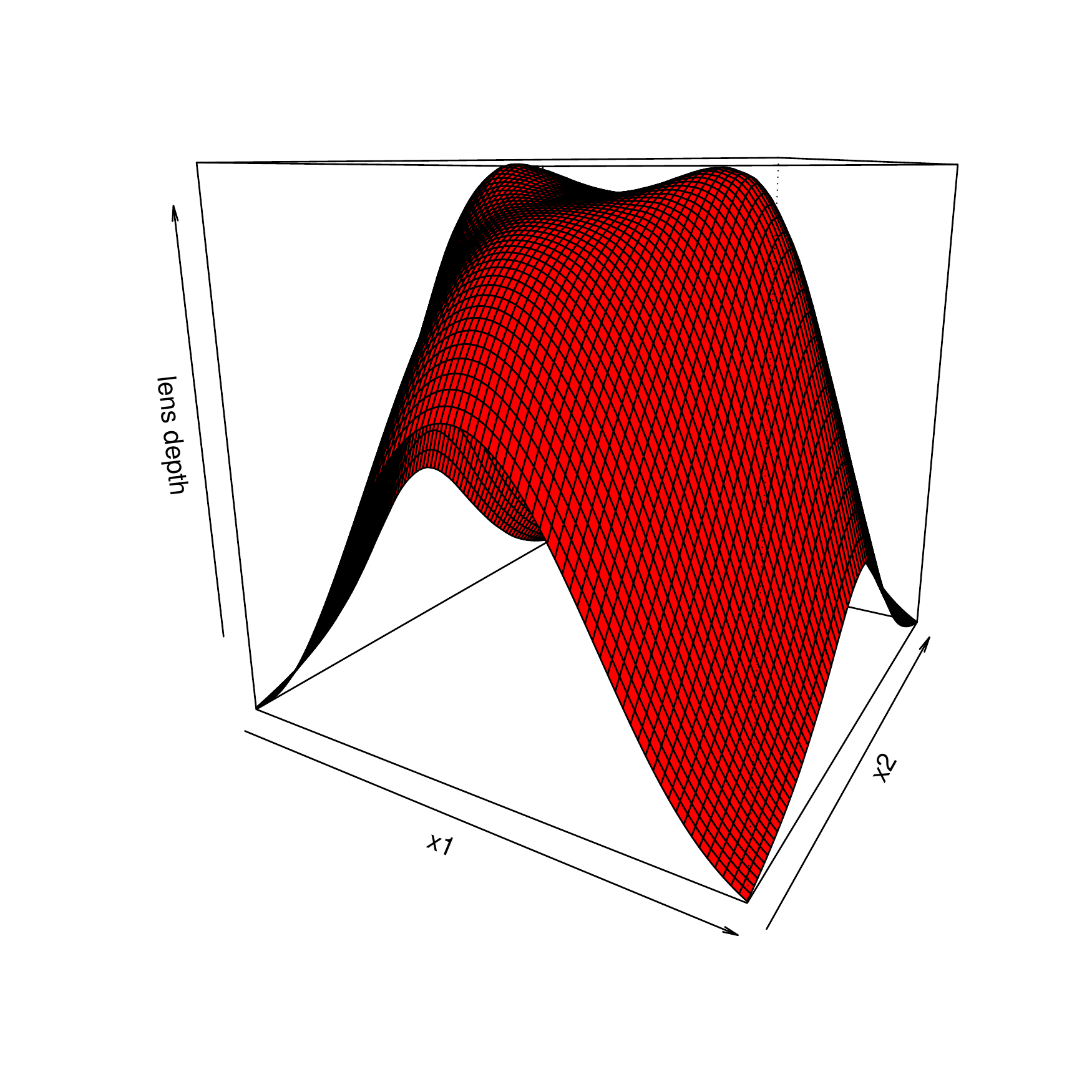}
	\includegraphics[width=0.2425\linewidth]{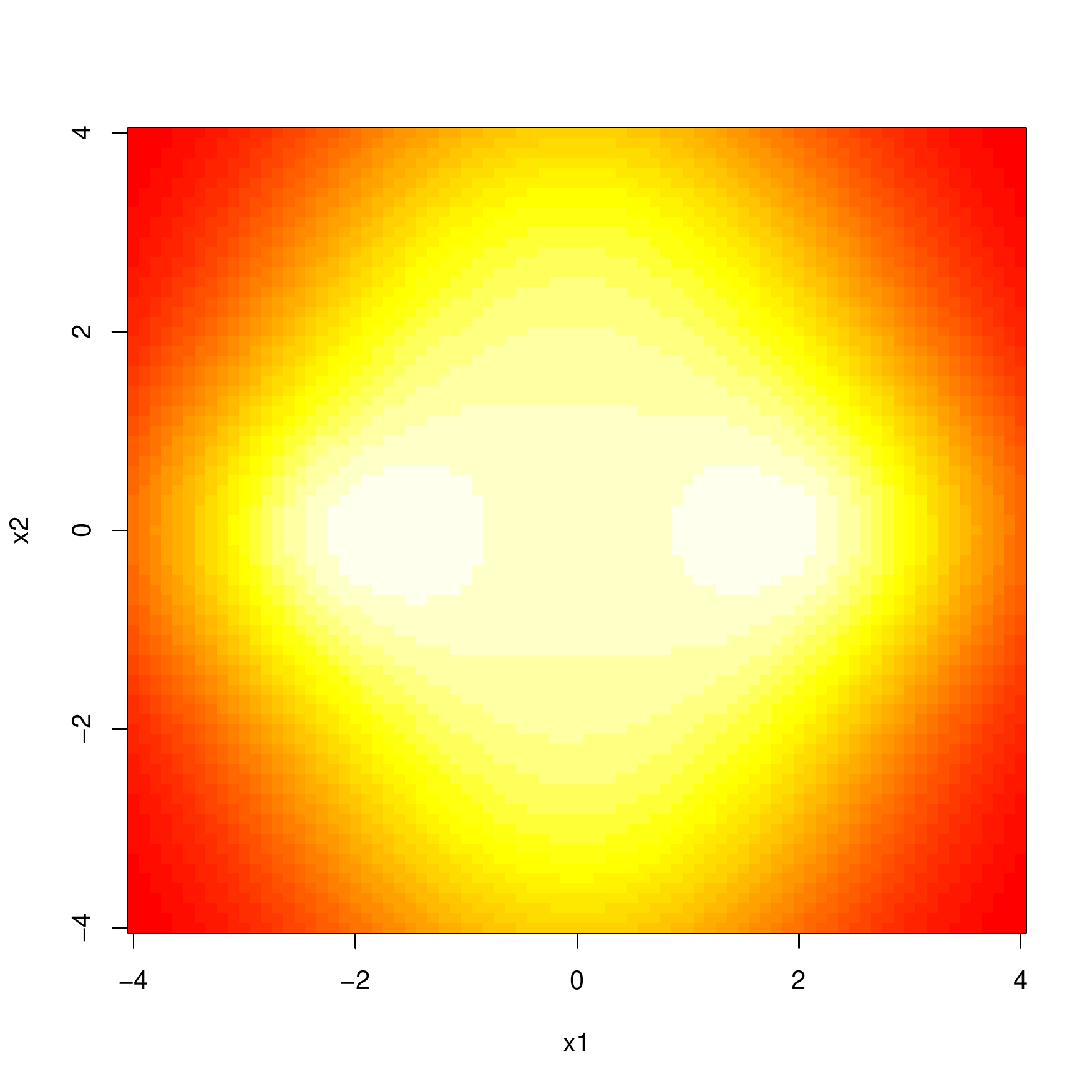}
	\includegraphics[width=0.2425\linewidth]{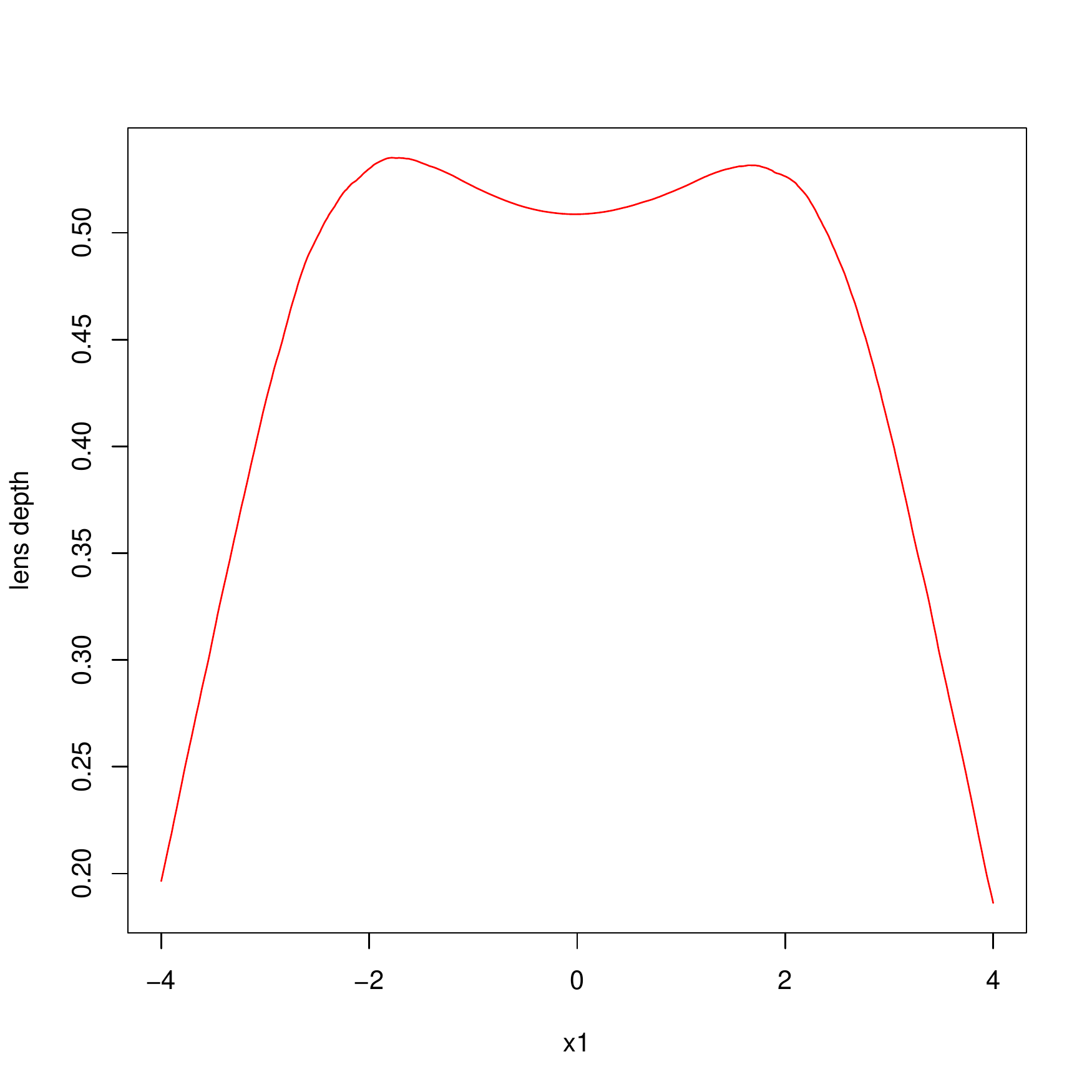}
	\includegraphics[width=0.2425\linewidth]{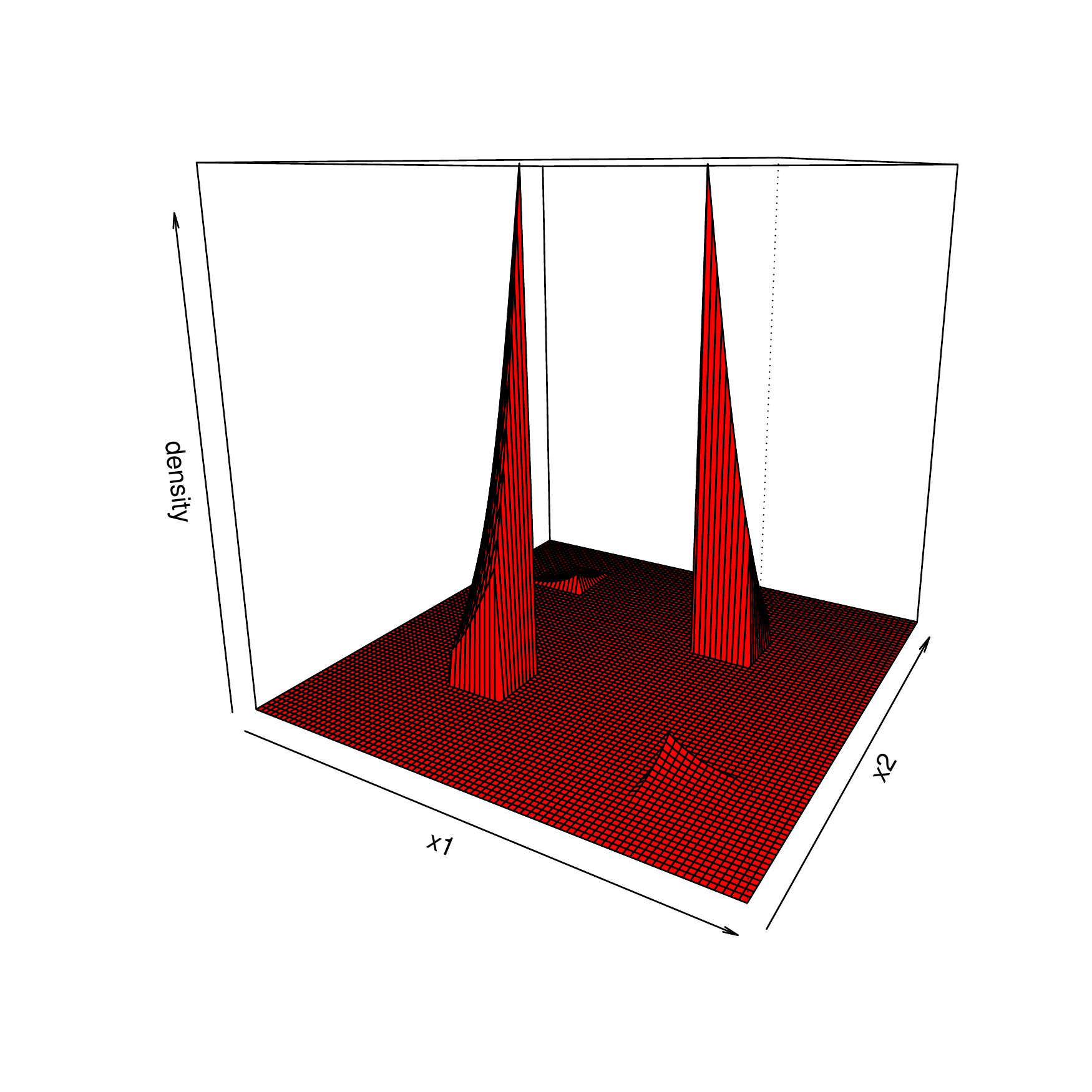}
	\includegraphics[width=0.2425\linewidth]{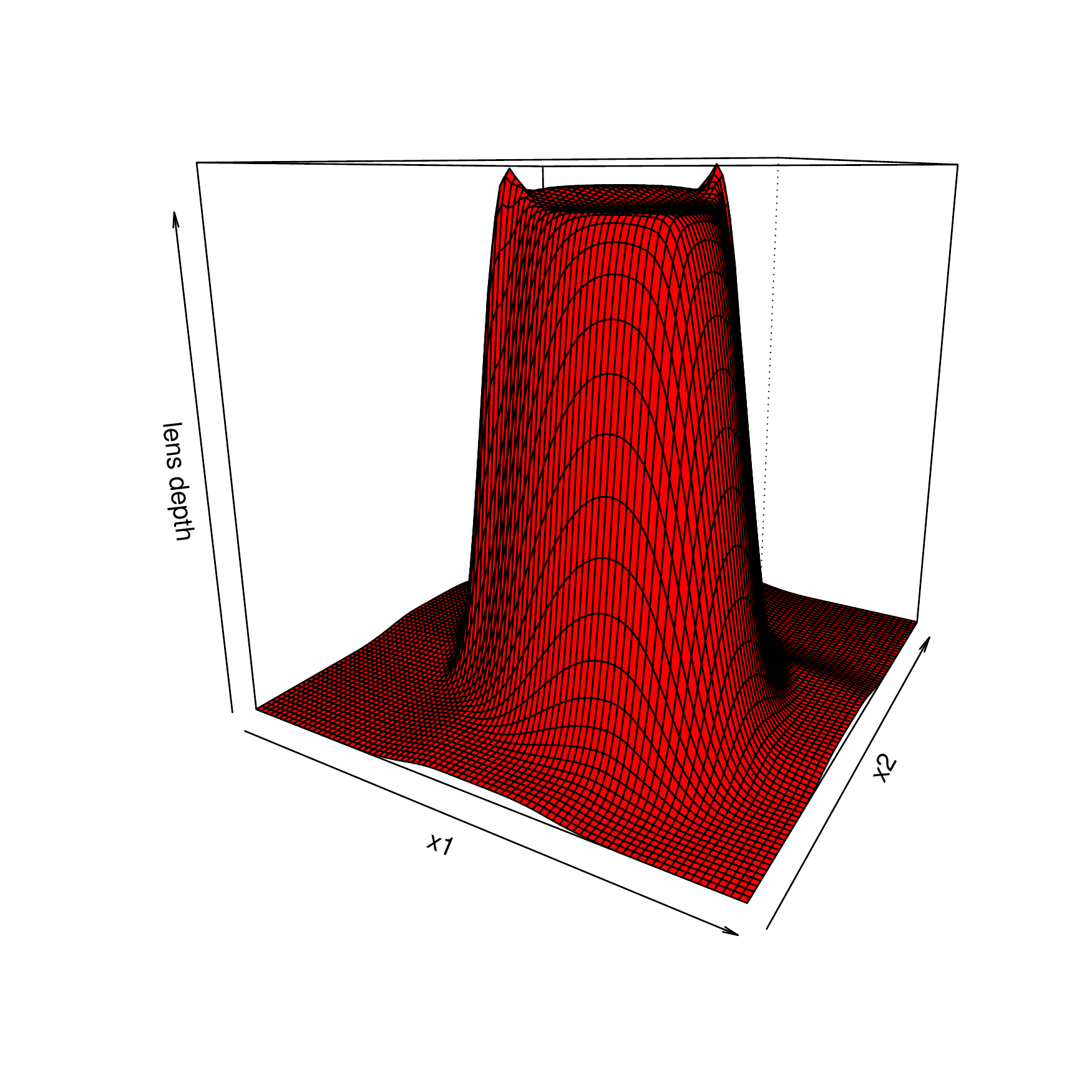}
	\includegraphics[width=0.2425\linewidth]{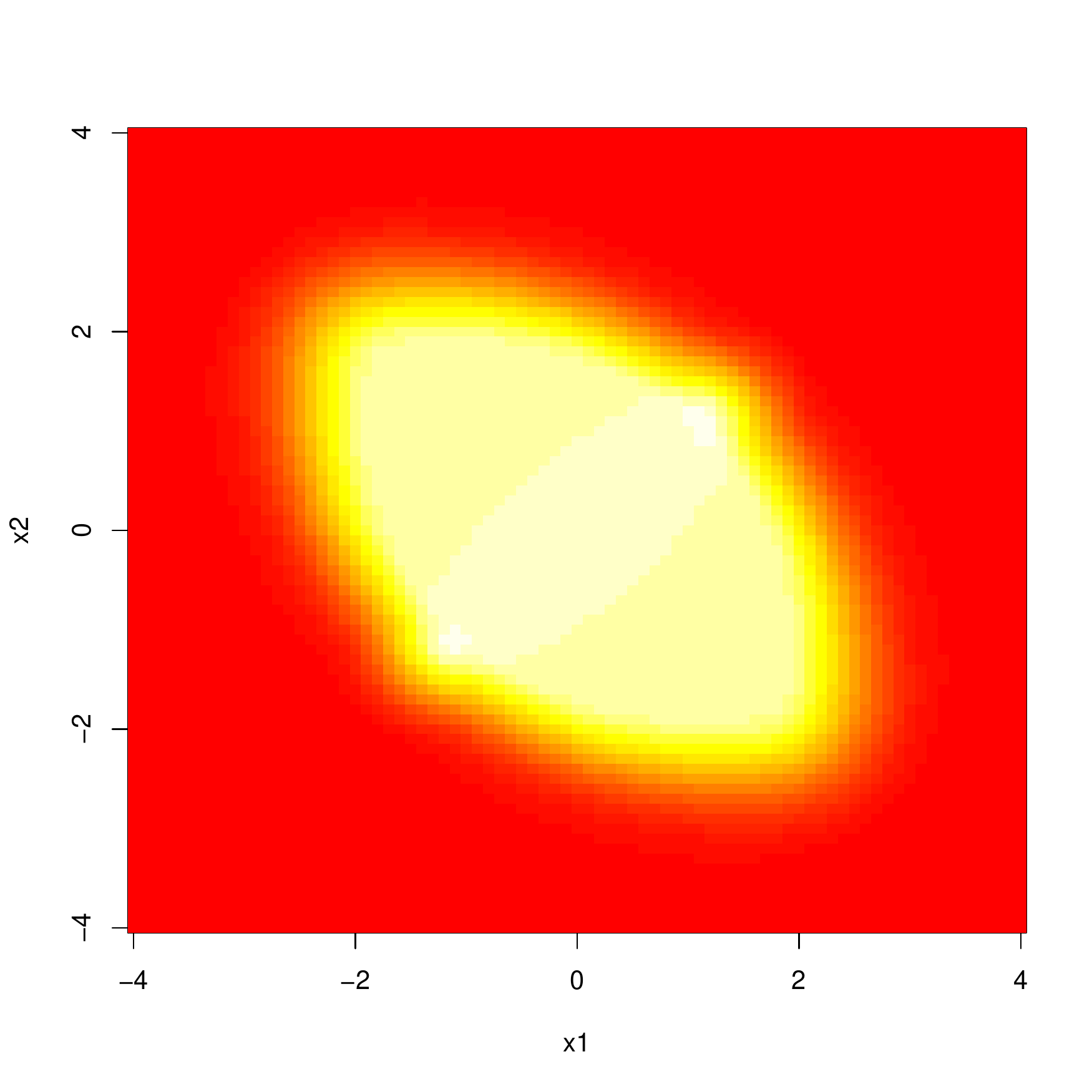}
	\includegraphics[width=0.2425\linewidth]{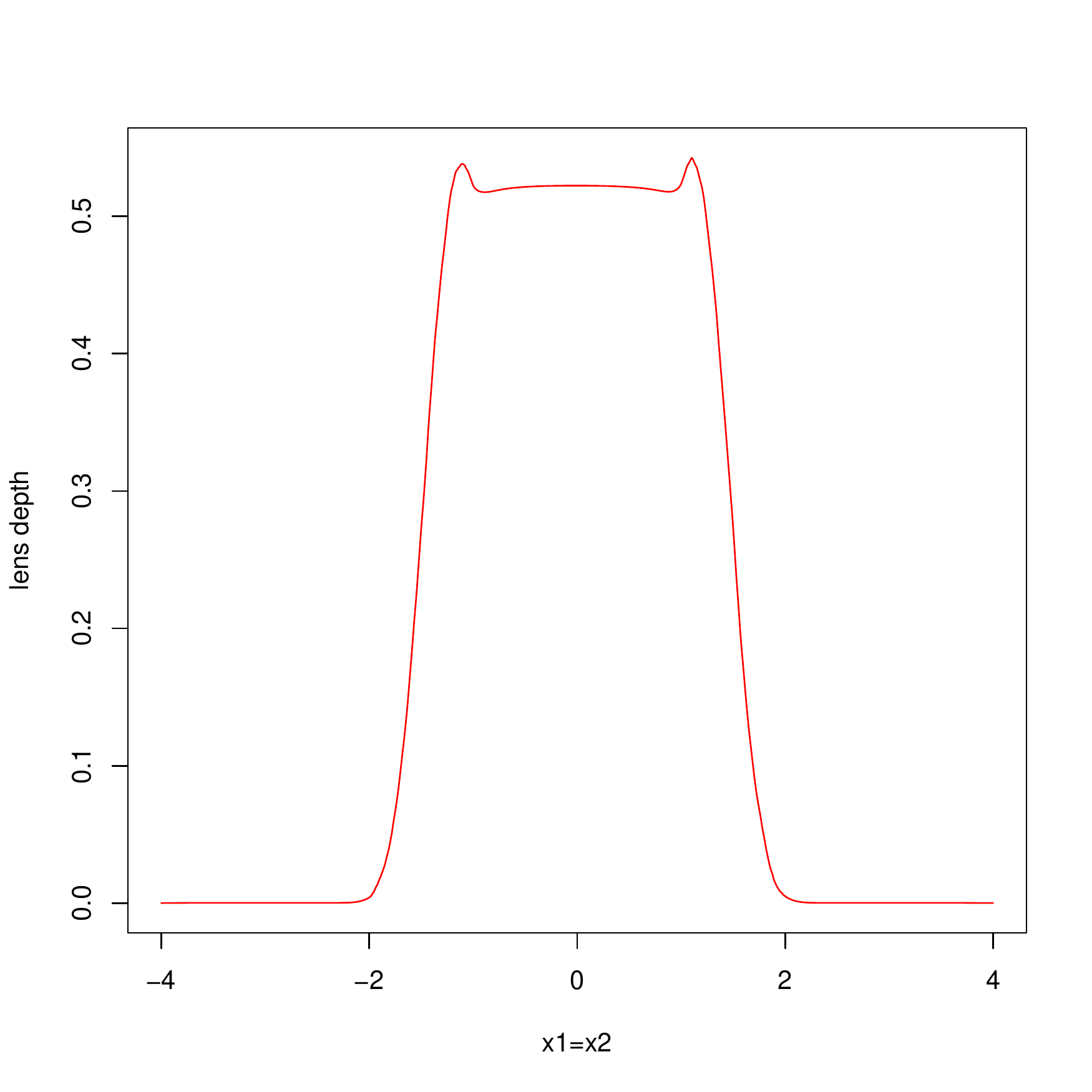}
	\includegraphics[width=0.2425\linewidth]{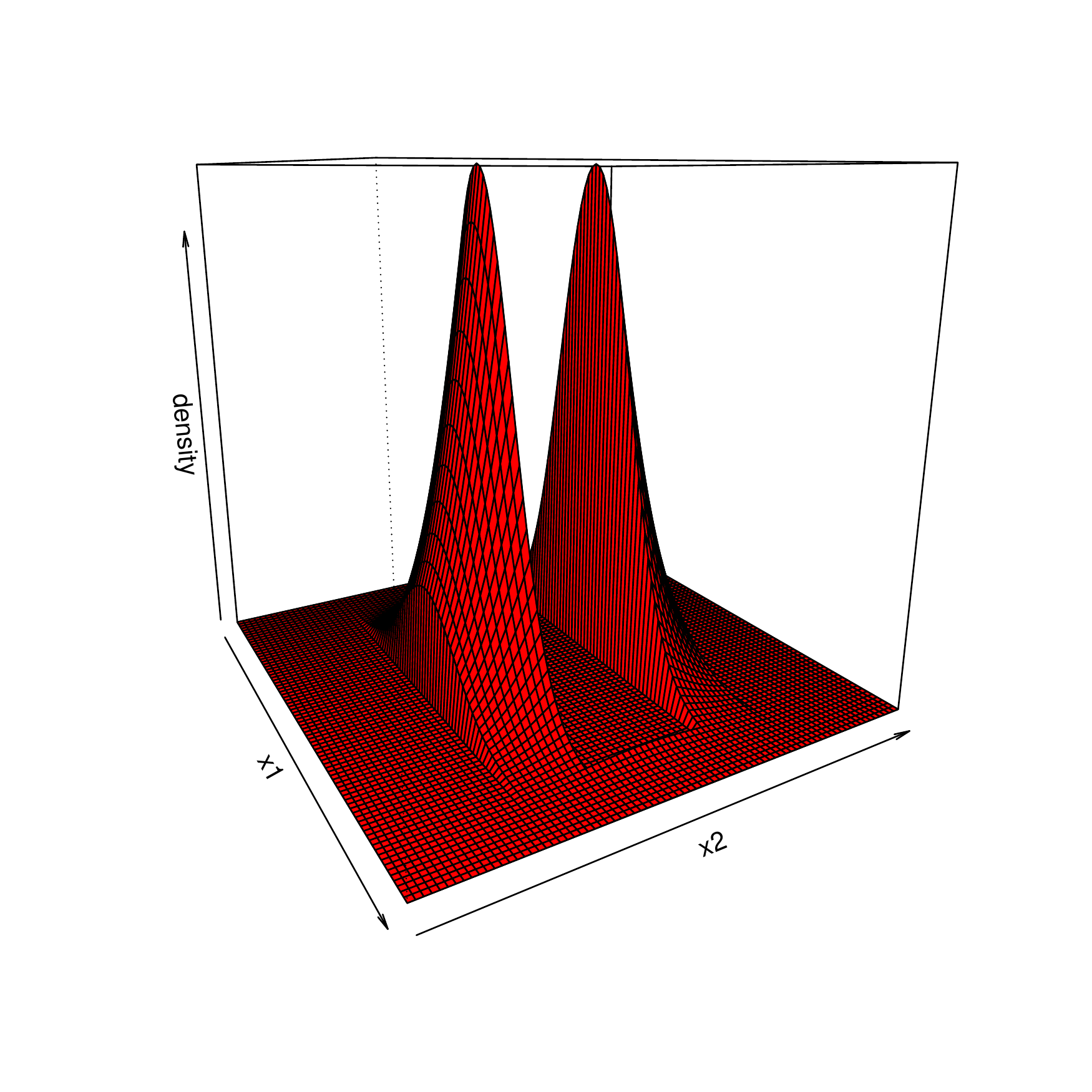}
	\includegraphics[width=0.2425\linewidth]{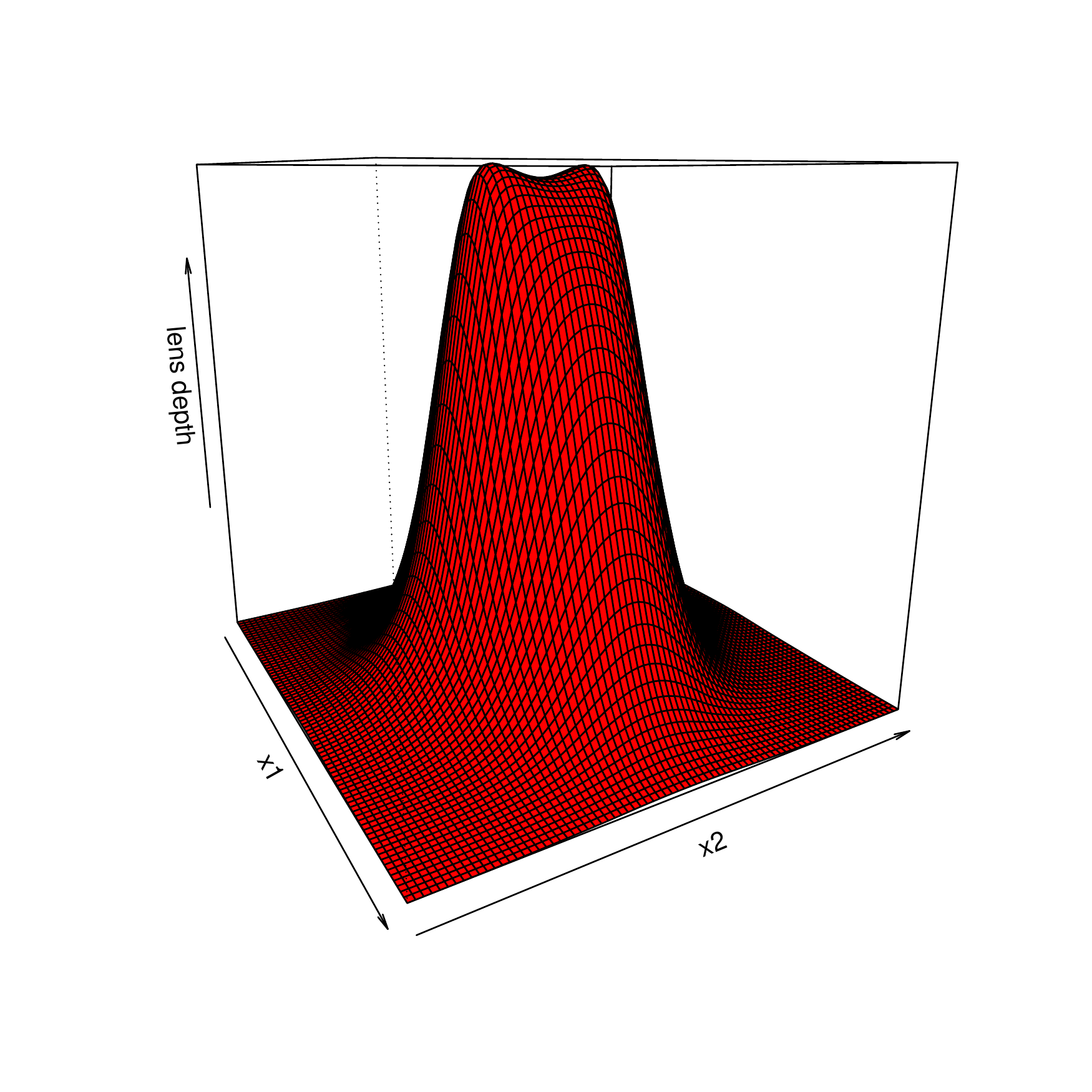}
	\includegraphics[width=0.2425\linewidth]{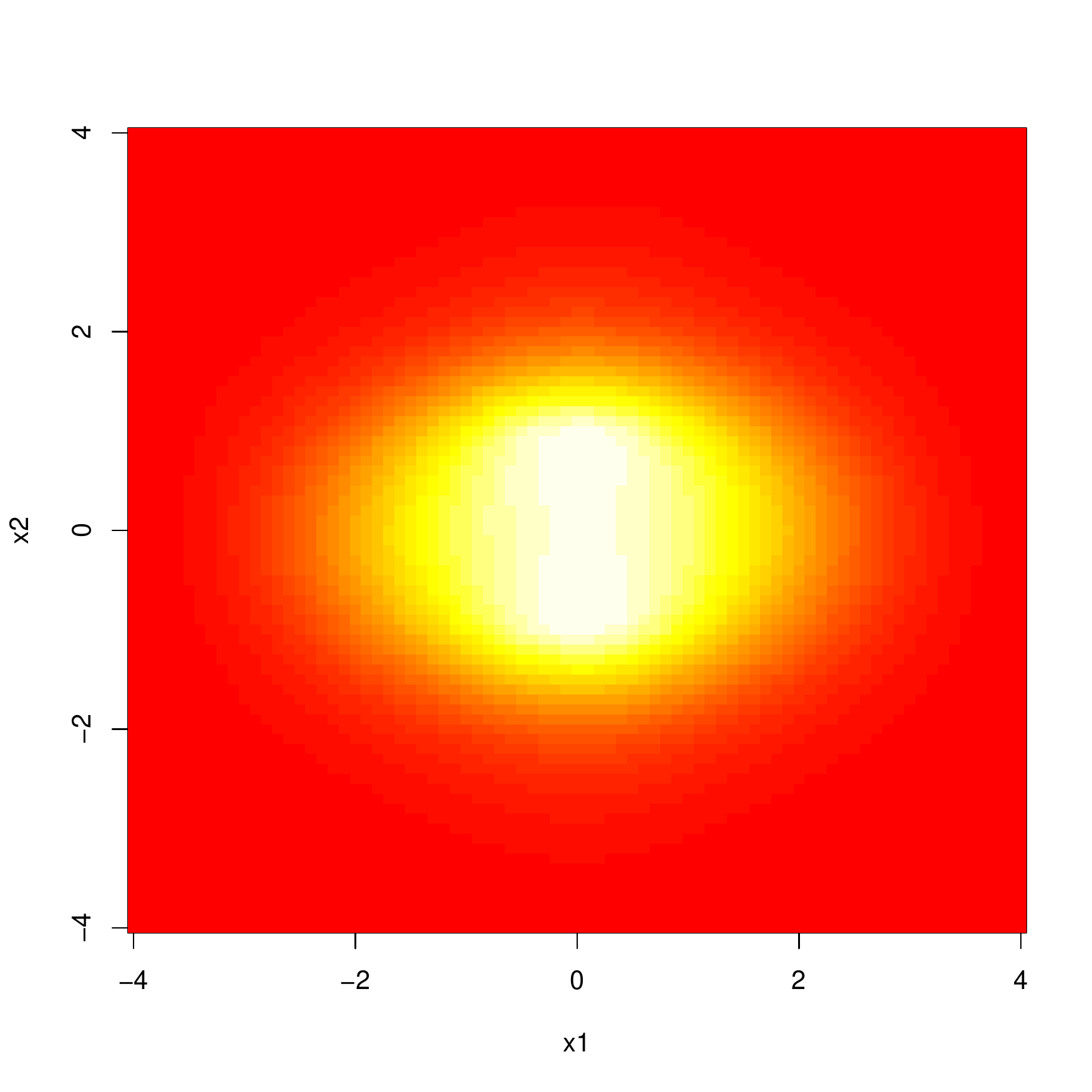}
	\includegraphics[width=0.2425\linewidth]{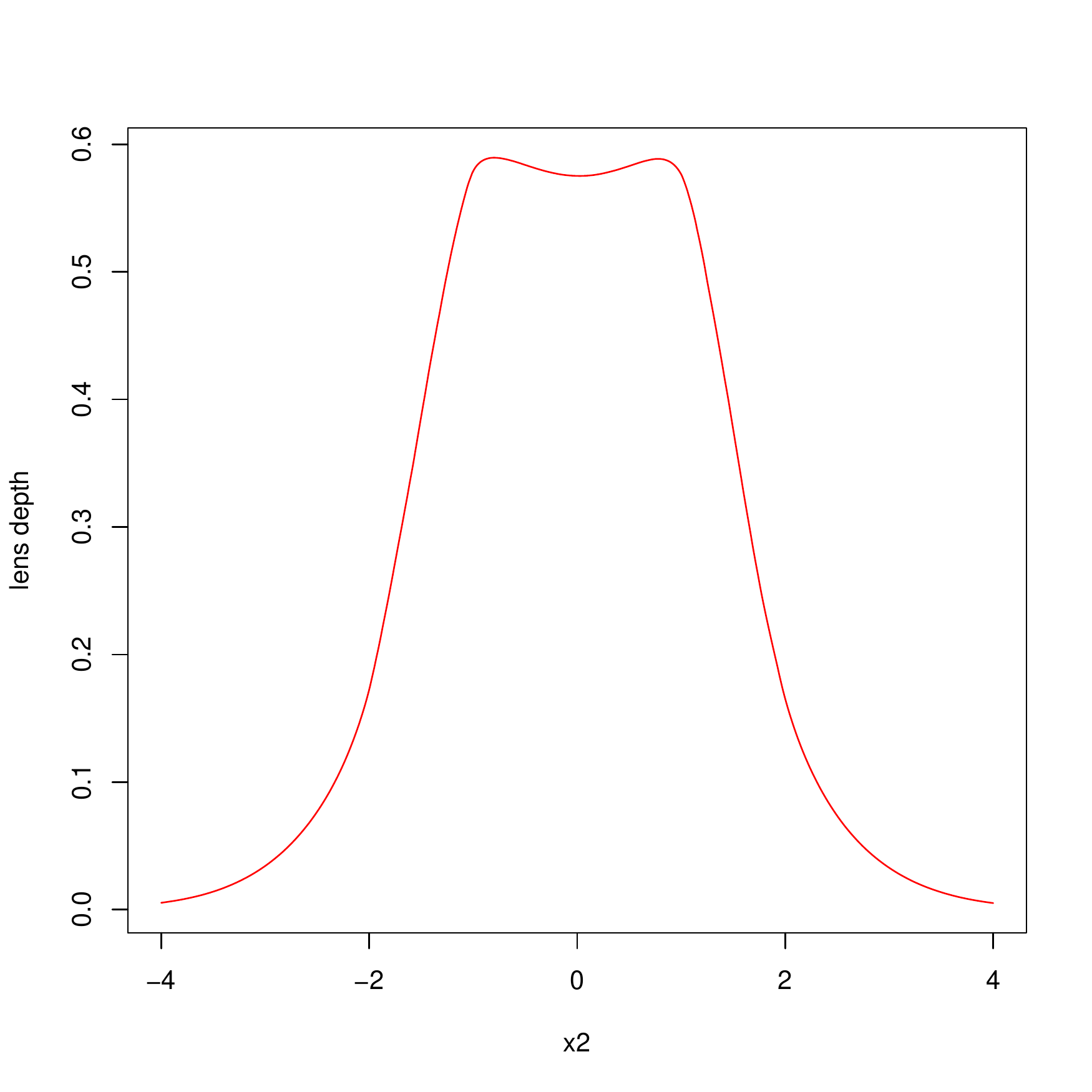}
	\caption{Top row: Example \ref{density_function_mixture_normals}; Central row: Example \ref{density_function_normal_truncated_to_squares}; Bottom row: Example  \ref{density_function_normal_truncated_to_frame}. From left to right, density function (first column),  sample lens depth constructed with $n=5,000$ sample draws from $P$ (second column), corresponding heat-map (third column) and its section along the line $x_{2}=0$ (top-right panel),  $x_{1}=x_{2}$ (central-right panel) and $x_{1}=0$ (bottom-right panel).}
	\label{figure:counterexample1}
\end{figure}

\begin{example}[Mixture of two normal distributions]\label{density_function_mixture_normals}
	Let $P$ be a mixture of two bivariate normal distributions with respective means $(-3,0)$, $(3,0)$, identity covariance matrices and equal weights -- viz., the density of $P$ is $f(x_1,x_2) = \frac{1}{2}\phi(x_1-3,x_2)+\frac{1}{2}\phi(x_1+3,x_2)$, for $\phi$ the standard bivariate normal density.
\end{example}
\begin{example}[Bivariate normal distribution truncated to 4 squares]\label{density_function_normal_truncated_to_squares}
	Let $P$ be the distribution 
	whose density function is
	$f(x_1,x_2)=\phi(x_1,x_2) / \left( \int_A \phi(x_1,x_2) \, dx_1 dx_2 \right)\indic{(x_1,x_2) \in A}, 
	$ 
	where $\phi$ is the standard bivariate normal density,  $A \doteq A_1 \cup A_2 \cup A_3 \cup A_4$ with $A_1=[-2,-1]\times[-2,-1]$, $A_2=[-3,-2]\times[2,3]$, $A_3=[1,2]\times[1,2]$ and $A_4=[2,3]\times[-3,-2]$. 
\end{example}
\begin{example}[Bivariate normal distribution truncated to a frame]\label{density_function_normal_truncated_to_frame}
	Let $P$ be as in Example \ref{density_function_normal_truncated_to_squares} but with $A_1=[-4,-3]\times[-1,1]$, $A_2=[-4,4]\times[1,2]$, $A_3=[3,4]\times[-1,1]$ and $A_4=[-4,4]\times[-2,-1]$. 
\end{example}
\ppn The distribution $P$ is clearly centrally symmetric about $(0,0)$ in each case: in Example \ref{density_function_mixture_normals} as $f(-x_1,-x_2)=f(x_1,x_2)$, and in Examples \ref{density_function_normal_truncated_to_squares}-\ref{density_function_normal_truncated_to_frame} because the standard bivariate normal distribution is centrally symmetric about $(0,0)$ and the region $A$ is symmetric with respect to the origin.
However, Figure \ref{figure:counterexample1} reveals that \cite{Liu11}'s Euclidean lens depth function with respect to any of these three distributions is not maximum at the centre $(0,0)$, nor is monotonic away from the deepest point(s) -- see that the depth function admits a local maximum at $(0,0)$ for Example \ref{density_function_normal_truncated_to_squares}.

\end{document}